\documentclass[aps,preprint]{revtex4}
\usepackage{amsmath}
\usepackage[english]{babel}
\usepackage{amssymb}
\usepackage{graphicx}
\usepackage{epsfig}
\usepackage{bm}
\usepackage{hyperref}
\usepackage{color}
\DeclareMathOperator{\sech}{sech}

\def\trmin{{\rm tr}}

\DeclareMathOperator*{\sumint}{%
\mathchoice%
  {\ooalign{$\displaystyle\sum$\cr\hidewidth$\displaystyle\int$\hidewidth\cr}}
  {\ooalign{\raisebox{.14\height}{\scalebox{.7}{$\textstyle\sum$}}\cr\hidewidth$\textstyle\int$\hidewidth\cr}}
  {\ooalign{\raisebox{.2\height}{\scalebox{.6}{$\scriptstyle\sum$}}\cr$\scriptstyle\int$\cr}}
  {\ooalign{\raisebox{.2\height}{\scalebox{.6}{$\scriptstyle\sum$}}\cr$\scriptstyle\int$\cr}}
}

\begin{document}

\renewcommand{\thefootnote}{\arabic{footnote}}
\setcounter{footnote}{0}

\title{\sc\Large{Diquark and nucleons under strong magnetic fields in the NJL model} \vspace*{0.5cm}}

\author{M. Coppola$^{a,b}$, D. Gomez Dumm$^{c}$ and N.N.\ Scoccola$^{a,b}$ \vspace*{0.1cm}}

\affiliation{$^{a}$ CONICET, Rivadavia 1917, 1033 Buenos Aires, Argentina}
\affiliation{$^{b}$ Physics Department, Comisi\'{o}n Nacional de Energ\'{\i}a At\'{o}mica, }
\affiliation{Avenue Libertador 8250, 1429 Buenos Aires, Argentina}
\affiliation{$^{c}$ IFLP, CONICET $-$ Departamento de F\'{\i}sica, Facultad de Ciencias Exactas,
Universidad Nacional de La Plata, C.C. 67, 1900 La Plata, Argentina \vspace*{2cm}}

\begin{abstract}
We study the description of nucleons and diquarks in the presence of a
uniform strong magnetic field within the framework of the two-flavor
Nambu-Jona--Lasinio (NJL) model. Diquarks are constructed through the
resummation of quark loop chains using the random phase approximation, while
nucleons are treated as bound quark-diquark states described by a
relativistic Fadeev equation, using the static approximation for  
quark exchange interactions.
For charged particles, analytical calculations are performed
using the Ritus eigenfunction method, which properly takes into account the
breakdown of translation invariance that arises from the presence of
Schwinger phases. Within this scheme, for definite model parametrizations we
obtain numerical predictions for diquark and nucleon masses, which are
compared with Chiral Perturbation Theory and Lattice QCD results. In
addition, numerical estimations for nucleon magnetic moments are obtained.
\end{abstract}
%\date{\today}

%\pacs{}%21.65.Qr, 25.75.Nq, 75.30.Kz, 11.30.Rd}

\maketitle

\renewcommand{\thefootnote}{\arabic{footnote}}
\setcounter{footnote}{0}

\section{Introduction}

In recent years a significant effort has been devoted to the study of the
properties of strongly interacting matter under the influence of strong
magnetic fields (see
e.g.~\cite{Kharzeev:2012ph,Andersen:2014xxa,Miransky:2015ava} and
refs.~therein). This is mostly motivated by the realization that large
magnetic fields might play an important role in the physics of the early
Universe~\cite{Grasso:2000wj}, in the analysis of high energy non-central
heavy ion collisions~\cite{HIC} and in the description of physical systems
such as magnetars~\cite{duncan}. From the theoretical point of view,
addressing this subject requires to deal with quantum chromodynamics (QCD)
in nonperturbative regimes. Therefore, existing analyses are based either in
the predictions of effective models or in the results obtained through
lattice QCD (LQCD) calculations. Most of these works have been focused on
the properties of light mesons. To deal with low energy QCD, various
theoretical approaches have been followed, e.g.\ Nambu-Jona-Lasinio
(NJL)-like
models~\cite{Fayazbakhsh:2013cha,Fayazbakhsh:2012vr,Avancini:2015ady,Zhang:2016qrl,Avancini:2016fgq,
Mao:2017wmq,GomezDumm:2017jij,Wang:2017vtn,Liu:2018zag,Coppola:2018vkw,Coppola:2019uyr,Mao:2018dqe,Avancini:2018svs},
quark-meson models~\cite{Kamikado:2013pya,Ayala:2018zat}, chiral
perturbation theory
(ChPT)~\cite{Andersen:2012zc,Agasian:2001ym,Colucci:2013zoa}, path integral
Hamiltonians~\cite{Orlovsky:2013wjd,Andreichikov:2016ayj}, Effective Chiral
Confinement Lagrangian approaches
(ECCL)~\cite{Simonov:2015xta,Andreichikov:2018wrc} and QCD sum rules
(SRQCD)~\cite{Dominguez:2018njv}. In addition, results for the light meson
spectrum in the presence of background magnetic fields have been obtained
from LQCD
calculations~\cite{Bali:2011qj,Hidaka:2012mz,Luschevskaya:2014lga,Bali:2015vua,Bali:2017ian,Ding:2020hxw}.
Regarding the study of other hadrons, in the last few years some works have
analyzed the effects of a magnetic field on baryon masses. This problem has
been addressed in the context of
ChPT~\cite{Tiburzi:2008ma,Deshmukh:2017ciw}, nonrelativistic quark
models~\cite{Taya:2014nha}, extended linear sigma
model~\cite{Haber:2014ula}, Walecka
model~\cite{Haber:2014ula,Mukherjee:2018ebw}, soliton
models~\cite{He:2016oqk}, Finite Energy QCD sum
rules~\cite{Dominguez:2020sdf}, and also lattice QCD~\cite{Endrodi:2019whh}.
It is worth noticing that these theoretical approaches lead to various
different results for the behavior of nucleon masses. The main purpose of
the present article is to complement these works by studying the effect of
an intense external magnetic field on scalar diquark and nucleon properties
within the NJL model.

In the framework of the NJL model, mesons and diquarks are usually described
as quantum fluctuations in the random phase approximation
(RPA)~\cite{Vogl:1991qt,Klevansky:1992qe,Hatsuda:1994pi}, i.e., they are
introduced via the summation of an infinite number of quark loops. In the
presence of a magnetic field $\vec B$, the calculation of these loops
requires some care due to the appearance of Schwinger
phases~\cite{Schwinger:1951nm} associated with quark propagators. For
neutral mesons Schwinger phases cancel out, and as a consequence one can
take the usual momentum basis to diagonalize the corresponding polarization
functions~\cite{Fayazbakhsh:2013cha,Fayazbakhsh:2012vr,Avancini:2015ady,Avancini:2016fgq,Mao:2017wmq}.
On the other hand, for charged pions and diquarks the Schwinger phases do
not cancel, leading to a breakdown of translational invariance that prevents
to proceed as e.g.~in the $\pi^0$ case. In this situation, some existing
calculations~\cite{Zhang:2016qrl,Liu:2018zag} just neglect Schwinger phases,
considering only the translational invariant part of the quark propagators.
Recently~\cite{Coppola:2018vkw,Coppola:2019uyr}, we have introduced a method
that allows to fully take into account the translational-breaking effects
introduced by the Schwinger phases in the calculation of charged meson
masses within the RPA. This method, based on the Ritus eigenfunction
approach~\cite{Ritus:1978cj} to magnetized relativistic systems, allows to
diagonalize the charged pion polarization function in order to obtain the
corresponding meson masses. In addition, in
Ref.~\cite{Coppola:2018vkw,Coppola:2019uyr} we have used a regularization
procedure in which only the vacuum contributions to different quantities at
zero external magnetic field are regularized. This scheme, that goes under
the name of ``Magnetic Field Independent Regularization'', has been shown to
provide more reliable predictions in comparison with other regularization
methods often used in the literature~\cite{Avancini:2019wed}. One of the
aims of the present work is to extend the Ritus eigenfunction approach to
the case of scalar diquarks. For this purpose we consider an extended
version of the NJL model that includes color pairing interactions.

As mentioned above, another aim of this work is to study the effects of an
external magnetic field on nucleon masses.  As shown some years
ago~\cite{Cahill:1988dx,Reinhardt:1989rw}, the quark level NJL
Lagrangian can be rewritten in terms of mesonic and baryonic degrees of
freedom, using diquarks as effective states in an intermediate step. As a
result of the hadronization process, one gets a relativistic Fadeev equation
that explicitly takes into account correlations among the three quarks. This
equation can be solved numerically in order to determine the nucleon
mass~\cite{Buck:1992wz,Huang:1993yd,Ishii:1993np,Ishii:1995bu}. In this way,
provided that the diquark channel interaction is strong enough, it is seen
that one can form a three-quark bound state with a phenomenologically
adequate nucleon mass. Using this framework, other nucleon properties have
been studied as well~\cite{Hellstern:1995ri, Asami:1995xq,Mineo:2002bg}. In
the present work we will follow this approach, considering the modifications
of the aforementioned Fadeev equation induced by the presence of an external
magnetic field. As expected, this leads to the existence of two different
Fadeev equations, one for the proton and another one for the neutron. Given
the complexity of the problem, we consider the static approximation
introduced in Ref.~\cite{Buck:1992wz}, which has been shown to lead to an
adequate description of nucleon properties in the absence of external
fields~\cite{Ishii:1993np}. Furthermore, for simplicity we neglect axial
vector diquark correlations.

This work is organized as follows. In Sec.~II we introduce the theoretical
formalism used to obtain the different quantities we are interested in. In
Sec.~III we present and discuss our numerical results. Finally, in Sec.~IV a
summary our work, together with our main conclusions, is given. We also
include Appendixes A and B to quote some technical details of our
calculations.

\section{Theoretical Formalism}

\subsection{Bosonized NJL model with diquark interactions in the presence of an external
magnetic field}

We start by considering the Euclidean Lagrangian density for the NJL
two-flavor model in the presence of an electromagnetic field and color pairing interactions. One has
\begin{equation}
{\cal L} \ =
\ \bar \psi \left(- i\, \gamma_\mu D_\mu + m_0 \right) \psi
- G \, \big[ j^{(S)}(x) j^{(S)}(x) + j^{(P)}_a(x) j^{(P)}_a(x) \big]
- H\, [j^{(D)}_A(x)]^\dagger j^{(D)}_A(x)\ ,
\label{lagrangian}
\end{equation}
where $\psi=(\psi_{u}\ \psi_{d})^T$, $G$ and $H$ are coupling constants, and
$m_0$ is the current quark mass, which is assumed to be equal for $u$ and
$d$ quarks. The currents in Eq.~(\ref{lagrangian}) are given by
\begin{eqnarray}
j^{(S)}(x) &=& \bar \psi(x) \, \psi(x) \ , \\
j^{(P)}_a(x) &=& \bar \psi(x)\, i \gamma_5 \, \tau_a \,\psi(x) \ , \\
j^{(D)}_A(x) &=& \bar \psi_c(x)\, i \gamma_5 \, \tau_2 \, \lambda_A \, \psi(x) \ ,
\end{eqnarray}
where we have defined $\psi_c = \gamma_2 \gamma_4 \bar \psi^T$, while
$\tau_a$ and $\lambda_A$, with $a=1,2,3$ and $A=2,5,7$, stand for Pauli and
Gell-Mann matrices acting on flavor and color spaces, respectively.

The interaction between the fermions and the electromagnetic field ${\cal
A}_\mu$ is driven by the covariant derivative
\begin{equation}
D_\mu \, = \, \partial_{\mu}-i\,\hat Q \mathcal{A}_{\mu}\ ,
\label{covdev}
\end{equation}
where $\hat Q=\mbox{diag}(Q_u,Q_d)$, with $Q_u=2e/3$ and $Q_d = -e/3$, $e$
being the proton electric charge. We consider the particular case of a
homogenous stationary magnetic field $\vec B$ orientated along the 3-axis.
Let us choose the Landau gauge, in which $\mathcal{A}_4 = 0$,
$\vec{\mathcal{A}} = (0,B x_1,0)$.

To proceed, it is convenient to bosonize the fermionic theory, introducing a
scalar field $\sigma(x)$, pseudoscalar fields $\vec{\pi}_a(x)$ and diquark
fields $\Delta_A(x)$, and integrating out the fermion fields. The bosonized
Euclidean action can be written as
\begin{equation}
S_{\mathrm{bos}} \, = \, - \frac12 \log\det \mathbf{D}+\frac{1}{4G}
\int d^{4}x\
\Big[\sigma(x)\,\sigma(x)+ \pi_a(x)\,\pi_a(x)  \Big]+ \frac{1}{4H}
\int d^{4}x \ \Delta_A(x)^\ast \Delta_A(x) \ ,
\label{sbos}
\end{equation}
where
\begin{equation}
 \mathbf{D}(x,x') \, =
\, \delta^{(4)}(x-x')\,
\left(%
\begin{array}{cc}
-i\gamma_\mu D_\mu  + m_0 + \phi(x)  &
 i\,\gamma_5\,\tau_2\, \lambda_A \, \Delta_A(x)\\
 i\,\gamma_5\,\tau_2\, \lambda_A \, \Delta_A(x)^\ast &
-i\gamma_\mu D^{\ast}_\mu  + m_0 + \phi(x)^T  \\
\end{array}%
\right)
\ ,
\label{dxx}
\end{equation}
with $\phi(x) = \sigma(x) + i\,\gamma_5\,\tau_a\pi_a(x)$. As customary, we
have used here the Nambu-Gorkov (NG) formalism. In the former equations, and
in what follows, matrices in the NG space are denoted in boldface.

We proceed by expanding the bosonized action in powers of the fluctuations
$\delta\sigma(x)$, $\delta\pi_a(x)$ and $\delta\Delta_A(x)$ around the
corresponding mean field (MF) values. As usual, we assume that the field
$\sigma(x)$ has a nontrivial translational invariant MF value
$\bar{\sigma}$, while the vacuum expectation values of pseudoscalar and
diquark fields are zero. Then, one has
\begin{equation}
\mathbf{D}(x,x') \ = \ \mathbf{ \bar D}(x,x') + \delta \mathbf{D}(x,x')\ ,
\label{dxxp}
\end{equation}
where the MF piece reads
\begin{eqnarray}
\mathbf{ \bar D}(x,x') \, = \,
\left(%
\begin{array}{cc}
{\cal{ \bar D}}(x,x') & 0\\
0 & {\cal{ \bar D}}_c(x,x')
\end{array}%
\right) \, = \, \delta^{(4)} (x-x')
\left(%
\begin{array}{cc}
-i\gamma_\mu D_\mu  + M & 0\\
0 & -i\gamma_\mu {D_\mu}^\ast  + M
\end{array}%
\right)
\ .
\end{eqnarray}
Here $M$ denotes the quark effective mass, $M= m_0 + \bar \sigma$. The
fluctuation piece is given by
\begin{eqnarray}
\delta \mathbf{D}(x,x') = \delta^{(4)} (x-x')
\left(%
\begin{array}{cc}
\delta\phi(x)  &
 i\,\gamma_5\,\tau_2\, \lambda_A \, \delta\Delta_A(x)\\
 i\,\gamma_5\,\tau_2\, \lambda_A \, \delta\Delta_A(x)^\ast &
\delta\phi(x)^T \\
\end{array}%
\right)\ .
\end{eqnarray}

The MF operators ${\cal{ \bar D}}(x,x')$ and ${\cal{ \bar D}}_c(x,x')$ are
flavor diagonal, and their inverses correspond to quark MF propagators in
the presence of a magnetic field. One has
\begin{eqnarray}
{\mathcal{ \bar D}}^{-1}(x,x')
&=&
{\cal{ \bar S}}(x,x') =
\ {\rm diag}
\big(
\mathcal{ \bar S}^{u}(x,x')\,
,\,
\mathcal{ \bar S}^{d}(x,x')
\big)\ ,
\\
{\mathcal{ \bar D}}_c^{-1}(x,x')
&=&
{\cal{ \bar S}}_c(x,x') =
\ {\rm diag}
\big(
\mathcal{ \bar S}^{-u}(x,x')\,
,\,
\mathcal{ \bar S}^{-d}(x,x')
\big)\ ,
\end{eqnarray}
where the minus signs in front of the flavor indices $f=u$ or $d$ indicate
that the sign of the corresponding quark electric charge in the propagator
has to be reversed. As is well known, the explicit form of the quark
propagator in the presence of an external constant magnetic field can be
written in different ways~\cite{Andersen:2014xxa,Miransky:2015ava}. For
convenience we take the form in which $\mathcal{ \bar S}^{f}(x,x')$ is given
by a product of a phase factor and a translational invariant function,
namely
\begin{equation}
\mathcal{ \bar S}^{f}(x,x') \ = \ e^{i\Phi_f(x,x')}\,\int_{p_\perp\, p_\parallel} e^{i\, p\, (x-x')}\, \tilde S^f(p_\perp,p_\parallel) \
,
\label{sfx}
\end{equation}
where $\Phi_f(x,x')= Q_f B (x_1+x_1')(x_2-x_2')/2$ is the so-called
Schwinger phase. We have introduced here the following shorthand notation
for the integrals over two-dimensional momentum vectors,
\begin{equation}
\int_{p\,q\, \dots}\ \equiv \ \int \dfrac{d^2 p}{(2\pi)^2} \, \dfrac{d^2 q}{(2\pi)^2}\, \dots\ .
\label{notation1}
\end{equation}
We find it convenient to express $\tilde S^f(p_\perp,p_\parallel)$ in the
Schwinger form~\cite{Andersen:2014xxa,Miransky:2015ava}
\begin{equation}
\tilde S^f(p_\perp,p_\parallel) \, = \, \int_0^\infty \!d\tau\,
%\exp\big[-\tau \phi_f(\tau,p)\big]\,
e^{-\tau \phi_f(\tau,p)}\,
\left\{\big(M-p_\parallel \cdot \gamma_\parallel \big)
\, \left[1+i s_f \,\gamma_1 \gamma_2\, \tanh(\tau B_f)\right] -
\dfrac{p_\perp \cdot \gamma_\perp}{\cosh^2(\tau B_f)} \right\} \ ,
\label{sfp_schw}
\end{equation}
where we have used the following definitions. The ``perpendicular'' and
``parallel'' gamma matrices are collected in vectors $\gamma_\perp =
(\gamma_1,\gamma_2)$ and $\gamma_\parallel = (\gamma_3,\gamma_4)$ (note that
in our convention $\{\gamma_\mu,\gamma_\nu\}=-2 \delta_{\mu\nu}$).
Similarly, $p_\perp = (p_1,p_2)$ and $p_\parallel = (p_3,p_4)$. We have also
used the notation $s_f = {\rm sign} (Q_f B)$ and $B_f=|Q_fB|$. Finally, we
have defined
\begin{equation}
\phi_f(\tau,p)\ = \ M^2 + p_\parallel^2 +
\dfrac{\tanh(\tau B_f)}{\tau B_f}\; p_\perp^2\ .
\end{equation}
Notice that the integral in Eq.~(\ref{sfp_schw}) is divergent and has to be
properly regularized, as we discuss below.

Replacing the previous relations in the bosonized effective action and
expanding in powers of the meson fluctuations around the MF values, one gets
\begin{eqnarray}
S_{\mathrm{bos}} \ = \ S^{\mbox{\tiny MF}}_{\mathrm{bos}} \, + \,
S^{\,\mbox{\tiny quad}}_{\mathrm{bos}}\, + \,\dots
\end{eqnarray}
The expression of $S^{\mbox{\tiny MF}}_{\mathrm{bos}}$, together with those
of the mesonic contributions to $S^{\,\mbox{\tiny quad}}_{\mathrm{bos}}$,
are given in Eqs.~(10-12) of Ref.~\cite{Coppola:2019uyr}. In that paper,
both the procedure followed to obtain the regularized gap equation and the
expressions required to calculate various meson properties are discussed in
detail. In the present case, $S^{\,\mbox{\tiny quad}}_{\mathrm{bos}}$
includes an additional contribution that is quadratic in the diquark fields.
This contribution will be discussed in the next subsection.

\subsection{Diquark mass and propagator}

The diquark contribution to $S^{\,\mbox{\tiny quad}}_{\mathrm{bos}}$ is
given by
\begin{equation}
S^{\,\mbox{\tiny quad,diq}}_{\mathrm{bos}} \ =\
S^{\,\mbox{\tiny quad},\Delta}_{\mathrm{bos}} +
S^{\,\mbox{\tiny quad},\bar\Delta}_{\mathrm{bos}} \ =\ \dfrac{1}{2}\,
\sum\limits_{D=\Delta,\bar{\Delta}} \: \int d^4x \, d^4x'\ \delta D_A(x)^\ast
\ {\cal G}^{-1}_D(x,x') \
\delta D_A(x')\ , \label{actionquad}
\end{equation}
where
\begin{eqnarray}
{\cal G}^{-1}_D(x,x') \ = \ \dfrac{1}{4 H}\; \delta^{(4)}(x-x') -
J_D(x,x')\ .
\end{eqnarray}
The polarization functions read
\begin{eqnarray}
J_{\Delta} (x,x') &=& \trmin_D \bigg[
\mathcal{ \bar S}^{u}(x,x') \ \gamma_5 \ \mathcal{ \bar S}^{-d}(x',x) \ \gamma_5 \
+
\mathcal{ \bar S}^{d}(x,x') \ \gamma_5 \ \mathcal{ \bar S}^{-u}(x',x) \ \gamma_5 \
\bigg]\ ,  \label{jotas} \\[.2cm]
J_{\bar \Delta} (x,x') &=& \trmin_D \bigg[
\mathcal{ \bar S}^{-u}(x,x') \ \gamma_5 \ \mathcal{ \bar S}^{d}(x',x) \ \gamma_5 \
+
\mathcal{ \bar S}^{-d}(x,x') \ \gamma_5 \ \mathcal{ \bar S}^{u}(x',x) \ \gamma_5 \
\bigg]\ ,
 \label{jotasb}
\end{eqnarray}
where the trace is taken over Dirac space. As seen from its quark content,
$\Delta$ ($\bar{\Delta}$) corresponds to the diquark with charge
$Q_\Delta=e/3$ ($Q_{\bar{\Delta}}=-e/3$). Since $J_\Delta (x,x')=J_{\bar
\Delta} (x',x)$, both diquarks have the same mass, and we can proceed by
considering only the positively charged diquark $\Delta$.

Let us start by replacing in Eq.~(\ref{jotas}) the expression for the quark
propagators in Eq.~(\ref{sfx}). We get
\begin{eqnarray}
\!\!\! J_{\Delta}(x,x') &=&  e^{i\Phi_\Delta(x,x')}\, \int_{p_\perp\,p_\parallel\,v_\perp\,v_\parallel}
 e^{i v (x-x')} \nonumber \\
& & \times\ \trmin_D\Big[
\tilde S^u(p_\perp^+,p_\parallel^+) \, \gamma_5 \, \tilde S^{-d}(p_\perp^-,p_\parallel^-)  \, \gamma_5
+ \
\tilde S^d(p_\perp^+,p_\parallel^+)  \, \gamma_5 \, \tilde S^{-u}(p_\perp^-,p_\parallel^-) \, \gamma_5
\Big]\ ,
\label{J+q1}
\end{eqnarray}
where we have defined $p^\pm = p \pm v/2$. Here the phase $\Phi_\Delta$ is
given by
\begin{eqnarray}
\Phi_\Delta(x,x') &=& \Phi_u(x,x') + \Phi_{-d}(x',x) \nonumber \\
& = & \Phi_d(x,x') + \Phi_{-u}(x',x) \nonumber \\
& = & \frac{Q_\Delta B}{2} (x_1+x'_1) ( x_2-x_2')\ ,
\end{eqnarray}
i.e., there is no cancellation of Schwinger phases. Consequently, the
polarization function is not translational invariant and will not become
diagonal when transformed to the momentum basis. In this situation, as done
in Ref.~\cite{Coppola:2019uyr} for the case of charged pions, it is
convenient to expand the diquark field in terms of Ritus eigenfunctions. We
have
\begin{equation}
\delta \Delta_A(x) \ = \ \sumint_{\bar q}\ \mathbb{F}_{\bar q}^\Delta(x) \;
\delta \Delta_A(\bar q) \ ,
\label{Ritus}
\end{equation}
where we have used the shorthand notation
\begin{equation}
\bar q \equiv (\ell,q_2,q_\parallel)\ ,
\qquad\qquad \sumint_{\bar q}\ \equiv \ \dfrac{1}{2\pi}\sum_{\ell=0}^\infty
\int \frac{dq_2}{2\pi}\,\int_{q_\parallel}\ .
\label{notation2}
\end{equation}
Notice that the expansion includes a sum over discrete Landau levels. The
functions $\mathbb{F}_{\bar q}^\Delta$ are given by
\begin{equation}
\mathbb{F}_{\bar q}^\Delta(x) \ = \ N_\ell \, e^{i ( q_2 x_2 + q_3 x_3 + q_4 x_4)}
\, D_\ell\left(\sqrt{2B_\Delta}\,x_1-s_\Delta\sqrt{2B_\Delta^{-1}}\,q_2\right)\ ,
\label{Fq}
\end{equation}
where $D_\ell(x)$ are the cylindrical parabolic functions and
$N_\ell= (4\pi B_\Delta)^{1/4}/\sqrt{\ell!}\,$. As in
Eq.~(\ref{sfp_schw}), we use the notation $B_\Delta = |Q_{\Delta} B|$ and
$s_\Delta= \mathrm{sign}(Q_{\Delta} B)$. Replacing now in
Eq.~(\ref{actionquad}) we have
\begin{equation}
S^{\,\mbox{\tiny quad},\Delta}_{\mathrm{bos}}\ = \ \frac{1}{2}\,\sumint_{\bar q', \bar q}
\;\delta\Delta_A(\bar q)^\ast\;
{\cal G}_\Delta^{-1}(\bar q, \bar q')
\; \delta\Delta_A (\bar q') \ ,
\label{actionquadTPF}
\end{equation}
where
\begin{equation}
{\cal G}_\Delta^{-1}(\bar q, \bar q') \ = \ \frac{1}{4H}\, \hat \delta_{\bar q \bar q'} -
J_\Delta(\bar q, \bar q')\ ,
\end{equation}
with
\begin{equation}
\hat \delta_{\bar q \bar q'} \ = \ (2\pi)^4\,
\delta_{\ell\ell'}\, \delta(q_2-q_2')\, \delta(q_3-q_3')\, \delta(q_4-q_4')
\end{equation}
and
\begin{eqnarray}
J_\Delta(\bar q, \bar q') & = &  \int_{p_\perp\,p_\parallel\,v_\perp\,v_\parallel}
 \trmin_D\Big[
\tilde S^u(p_\perp^+,p_\parallel^+) \, \gamma_5 \, \tilde S^{-d}(p_\perp^-,p_\parallel^-)  \, \gamma_5
 +
\tilde S^d(p_\perp^+,p_\parallel^+)  \, \gamma_5 \, \tilde S^{-u}(p_\perp^-,p_\parallel^-) \, \gamma_5
\Big]
\nonumber \\
& & \qquad \qquad \times \ \int d^4x \, d^4x'\,
e^{i\Phi_\Delta(x,x')} e^{i v(x-x')} \,
\mathbb{F}_{\bar q}^\Delta(x)^\ast\,
\mathbb{F}_{\bar q'}^\Delta(x')\ .
\label{Fmu+1}
\end{eqnarray}
The integrals in Eq.~(\ref{Fmu+1}) can be worked out following basically the
same steps as those described in Ref.~\cite{Coppola:2019uyr} for the case of
charged pions. In this way, after some lengthy calculation, it can be
shown that the polarization function turns out to be diagonal in the Ritus
eigenfunction basis. One has
\begin{equation}
J_\Delta(\bar q, \bar q') \ = \ \hat \delta_{\bar q \bar q'} \; J_{\Delta}(\ell,\Pi^2)\ ,
\end{equation}
where
\begin{eqnarray}
J_{\Delta}(\ell,\Pi^2) &=& \dfrac{1}{2\pi^2} \int_0^\infty\! dz
\int_{0}^1 dy \ \exp\big[-zM^2-zy(1-y)(\Pi^2-(2\ell+1)\, B_\Delta)\big]
\nonumber \\
&& \times\,\dfrac{\alpha_-^\ell}{\alpha_+^{\ell+1}}\;
\bigg\{\Big[M^2+\dfrac{1}{z}-y(1-y)(\Pi^2-(2\ell+1)\, B_\Delta)
\Big](1+t_u \,t_d)
\nonumber \\
&& +\,\dfrac{(1-t_u^2)(1-t_d^2)}{\alpha_+\,\alpha_-}\, \Big[ \alpha_- +
(\alpha_- - \alpha_+)\,\ell \Big] \bigg\}\ ,
\label{J-B}
\end{eqnarray}
with $\Pi^2 = (2\ell+1)B_\Delta+q_\parallel^2$. Here we have introduced the
definitions $t_u=\tanh (B_u y z)$, $t_d=\tanh [B_d (1-y) z]$ and $\alpha_\pm
= (B_d t_u+B_u t_d \pm B_\Delta \,t_ut_d)/(B_u B_d)$. As usual, we have
introduced the changes of variables $y = \tau/(\tau +\tau')$ and $z = \tau +
\tau'$, $\tau$ and $\tau'$ being the integration parameters associated with
the quark propagators as in Eq.~(\ref{sfp_schw}).

As in the case of the mesons~\cite{Coppola:2018vkw,Coppola:2019uyr},
the polarization function in Eq.~(\ref{J-B}) turns out to be divergent and
can be regularized within the Magnetic Field Independent Regularization
scheme. Due to quantization in the 1-2 plane this requires some care,
viz.~the subtraction of the $B=0$ contribution to the polarization function
has to be carried out once the latter has been written in terms of the
squared canonical momentum $\Pi^2$, as in Eq.~(\ref{J-B}). Thus, the
regularized diquark polarization function can be written as
\begin{equation}
J_{\Delta}^{{\rm (reg)}}(\ell,\Pi^2) \ = \ J^{\rm
(reg)}_{\Delta,B=0}(\Pi^2) \, +  \, J_{\Delta}^{{\rm (mag)}}(\ell,\Pi^2)\
, \label{J-reg}
\end{equation}
where
\begin{eqnarray}
J_{\Delta}^{{\rm (mag)}}(\ell,\Pi^2) &=& \frac{1}{2\pi^2}
\int_0^\infty\! dz \int_{0}^1 dy \;\exp\big[-z M^2-z y(1-y)\Pi^2\big]
\nonumber \\
&& \times\,\bigg\{ \bigg[ M^2 +\dfrac{1}{z} - y(1-y)\Big[\Pi^2-(2\ell+1)\,B_\Delta\Big] \bigg]\nonumber \\
&& \times\,\bigg[\,\dfrac{\alpha_-^\ell}{\alpha_+^{\ell+1}}\,(1+t_u \,t_d)\,
\exp\big[z\,y(1-y)(2\ell+1)B_\Delta\big] - \dfrac{1}{z} \bigg]
\nonumber \\
&& +\,\dfrac{\alpha_-^{\ell-1}}{\alpha_+^{\ell+2}}\,(1-t_u^2)\,(1-t_d^2)\,
\Big[ \alpha_- + (\alpha_- - \alpha_+)\,\ell \Big]
\nonumber \\
&& \times\,\exp\big[z\,y(1-y)(2\ell+1)\,B_\Delta\big] - \dfrac{1}{z} \bigg[
\dfrac{1}{z} - y(1-y)(2\ell+1)\,B_\Delta \bigg] \bigg\}\ .
\label{J-mag}
\end{eqnarray}
The integrand in Eq.~(\ref{J-mag}) is well behaved in the limit $z\to 0$.
Hence, this magnetic field-dependent contribution is finite. On the other
hand, the expression for the subtracted $B=0$ piece has to be regularized.
This can be done, as usual, by using a 3D cutoff regularization. We get
\begin{equation}
J^{\rm (reg)}_{\Delta,B=0}(\Pi^2) =
2 \left[ I_1 + \Pi^2 I_2(\Pi^2) \right]\ ,
\end{equation}
where the explicit expressions of $I_1$ and $I_2$ can be found e.g.\ in
Ref.~\cite{Coppola:2019uyr} [see Eqs.~(20) and (28)]. We obtain in this way
\begin{equation}
{\cal G}^{-1}_\Delta(\bar q, \bar q')\ = \ \hat\delta_{\bar q \bar q'}
 \, \left[ \frac{1}{4H}\, - J_{\Delta}^{{\rm (reg)}}(\ell,\Pi^2) \right]\ .
\label{2pf}
\end{equation}
Since the two-point function is diagonal in this basis, it can be trivially inverted
to obtain the diquark propagator. We have
\begin{equation}
{\cal G}_\Delta(\bar q, \bar q') \ = \ \hat\delta_{\bar q \bar q'}
\; {\cal G}^{\rm (reg)}_\Delta(\ell,q_\parallel^2)\ ,
\label{diqprop}
\end{equation}
where
\begin{equation}
{\cal G}^{\rm (reg)}_\Delta(\ell,q_\parallel^2) \ = \
\left[ \frac{1}{4H}\, - J_{\Delta}^{{\rm (reg)}}(\ell,\Pi^2) \right]^{-1}\ .
\label{glq}
\end{equation}
Consequently, in our framework the diquark pole mass in the presence of the
magnetic field for each Landau level $\ell$ can be obtained by solving the
equation
\begin{equation}
\frac{1}{4H} - J_{\Delta}^{{\rm (reg)}}(\ell,-m_{\Delta}^2) \ = \ 0 \ .
\label{polediquark}
\end{equation}
It is clear that $m_\Delta$ depends on the magnetic field, although not
explicitly stated.

As in the case of the charged pions, instead of dealing with $m_{\Delta}$
one can define the $\Delta$ ``magnetic field-dependent mass'' as the lowest
quantum-mechanically allowed energy of the diquark, $E_\Delta$. The latter
is given by
\begin{equation}
E_{\Delta}^2 \ = \  m_{\Delta}^2 + (2\ell+1)\,B_\Delta + q_3^2\;\Big|_{q_3 =0,\,\ell=0}
\ = \ m_{\Delta}^2 + \frac{|eB|}{3}\ .
\label{epimas}
\end{equation}
Notice that this ``mass'' is magnetic field dependent even for a pointlike
diquark (in which case one would have a pole mass $m_\Delta$ independent of
$B$). In fact, owing to zero-point motion in the 1-2 plane, even for
$\ell=0$ a diquark cannot be at rest in the presence of the magnetic field.

Given the diagonal form of the diquark propagator in Ritus space, see
Eq.~(\ref{diqprop}), we can transform it back to coordinate space.
One obtains
\begin{equation}
{\cal G}_\Delta(x,x') \ = \ e^{i \Phi_\Delta(x,x')} \,
\int_{q_\perp \, q_\parallel} e^{i q (x-x')} \,  \tilde {\cal
G}_\Delta(q_\perp,q_\parallel)\ ,
\label{diqpro}
\end{equation}
where
\begin{equation}
\tilde {\cal G}_\Delta(q_\perp,q_\parallel) \ = \ 2 \,
e^{-q_\perp^2/B_\Delta} \ \sum_{\ell=0}^{\infty} \ (-1)^\ell \, {\cal G}^{\rm (reg)}_\Delta(\ell,q_\parallel^2)
\, L_\ell\left( 2 q_\perp^2/B_\Delta \right)\ ,
\label{diprop}
\end{equation}
$L_\ell(x)$ being the Laguerre polynomials.

\subsection{Nucleon masses}

\begin{figure}[htb]
    \centering{}\includegraphics[width=0.95\textwidth]{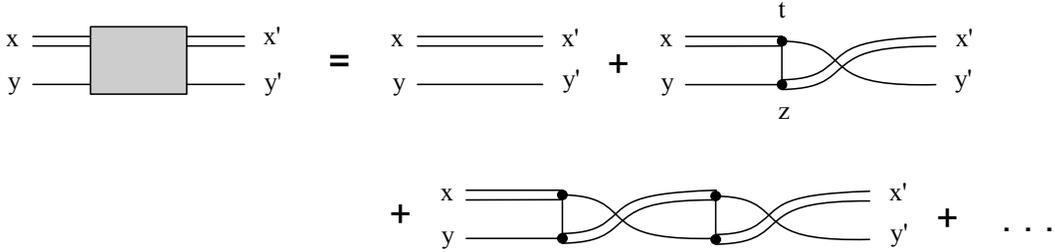}
\caption{Diagrams contributing to the full baryon propagator.}
\label{fig:1}
\end{figure}

The baryon propagator can be obtained consistently with the bound
quark-diquark structure following Ref.~\cite{Reinhardt:1989rw}. From the
infinite sum illustrated by the diagrams in Fig.~\ref{fig:1} one arrives
at a relation of the form
\begin{eqnarray}
{\cal S}^B([x;y],[x';y'])& = & {\cal S}_0^B([x;y],[x';y']) \, \nonumber \\
&& + \int d^4t\,d^4z\;
{\cal S}_0^B([x;y],[t;z])\,{\cal H}(z,t) \,{\cal S}_0^B([z;t],[x';y']) + \dots
%\nonumber \\
\label{infsum}
\end{eqnarray}
where, in our case, the kernel ${\cal H}$ is given by
\begin{equation}
 {\cal H}(z,t) \ = \ \,i\gamma_5\tau_2\lambda_{A}\,
{\cal{ \bar S}}_c(z,t)\,i\gamma_5\tau_2\lambda_{A'}\ .
\end{equation}
In Eq.~(\ref{infsum}), ${\cal S}^B$ stands for the full baryon propagator,
while ${\cal S}_0^B$ describes the unperturbed propagation of a diquark and
a quark, namely
\begin{equation}
{\cal S}^B_0([x;y],[t;z]) \ = \ {\cal G}_\Delta(x,t) \, {\cal \bar
S}(y,z)\ .
\end{equation}
Since the nucleon fields are bilocal, we have introduced the notation of
pairs $[x;y]$, where the first and second coordinates correspond to the
diquark and the quark, respectively. The resummation of the diagrams in
Fig.~\ref{fig:1} leads to a relativistic Fadeev equation that can be written
in the form
\begin{equation}
{\cal S}^B_0([x;y],[x';y']) \,=\, \int d^4t\,d^4z \Big[
\delta^{(4)}(x-z)\delta^{(4)}(y-t) - L([x;y],[z;t])\Big]\,{\cal
S}^B([z;t],[x';y'])\ ,
\label{fadeev}
\end{equation}
where
\begin{equation}
L([x;y],[z;t]) \ = \ {\cal S}^B_0([x;y],[t;z])\, {\cal H}(z,t)\ .
\end{equation}

The nucleon masses will be given by the poles of the baryon propagator in
the background of the vacuum configuration of the meson fields. These poles
correspond to the zeros of the operator in square brackets in
Eq.~(\ref{fadeev}). Acting on the baryon field $\psi$, one has
\begin{equation}
\int d^4z\; d^4t \ L([x;y],[z;t]) \, \psi([z;t]) \ = \ \psi([x;y])\ .
\label{baruno}
\end{equation}
It should be noticed that in our calculation only isocalar-scalar diquark
interactions have been considered. This implies that the nucleon isospin is
directly given by the flavor of the unpaired quark. Projecting on color
singlet baryon states, and using the explicit form of the matrices in flavor
space, one gets
\begin{eqnarray}
&& 2 \int d^4z \, d^4t \ {\cal G}_\Delta(x,t)\, {\cal{ \bar S}}^u(y,z) \, \gamma_5\,  {\cal \bar S}^{-d}(z,t) \,
\gamma_5\,\psi_p([z;t]) \ = \ \psi_p([x;y]) \ ,
\\
&& 2 \int d^4z \, d^4t \ {\cal G}_\Delta(x,t)\, {\cal{ \bar S}}^d(y,z) \, \gamma_5\,  {\cal \bar S}^{-u}(z,t) \,
\gamma_5\,  \psi_n([z;t]) \ = \ \psi_n([x;y]) \ ,
\label{bardos}
\end{eqnarray}
where $\psi_p$ and $\psi_n$ stand for the proton and neutron states,
respectively.

It should be noticed that in the absence of an external magnetic field both
equations coincide. Moreover, since in that case both the quark and diquark
fields are translational invariant, one can perform a Fourier transformation
into momentum space. The resulting Fadeev equation, discussed e.g.\ in
Refs.~\cite{Buck:1992wz,Ishii:1993np}, turns out to be a non-separable
integral equation. Given its complexity, in Ref.~\cite{Buck:1992wz} the
so-called ``static approximation'', in which one disregards the momentum
dependence of the exchanged quark, was used. Then, in
Ref.~\cite{Ishii:1993np} the full equation was solved numerically, showing
that in fact the static approximation can be taken as a good qualitative
approach to the exact results. Having this in mind, and taking into account
the additional difficulty introduced by the external magnetic field, we find
it appropriate to consider the static approximation to get an estimation of
the behavior of nucleon masses with the external field. This means to take
\begin{eqnarray}
\tilde S^{-f}(p_\perp,p_\parallel) \ \rightarrow \ \frac{1}{M} \ .
\end{eqnarray}
Since in this approximation one has ${\cal \bar S}^{-f}(x,y) =
\delta^{(4)}(x-y)$ and $\mathcal{H}(x,z)\propto \delta^{(4)}(x-z)$,
Eqs.~(\ref{bardos}) reduce to
\begin{eqnarray}
&&\frac{2}{M} \int d^4z  \; {\cal G}_\Delta(x,z)\, {\cal{ \bar S}}^u(x,z) \,
\psi_p(z) \ = \ \psi_p(x) \ ,
\nonumber \\
&&\frac{2}{M} \int d^4z \; {\cal G}_\Delta(x,z)\, {\cal{ \bar S}}^d(x,z) \,
\psi_n(z) \ = \ \psi_n(x)\ .
\label{bartres}
\end{eqnarray}
Notice that within this approximation there is no further need to consider
coordinate pairs in the arguments of nucleon fields, which become local.

Inserting Eqs.~(\ref{sfx}) and (\ref{diqpro}) into Eqs.~(\ref{bartres}), we get
\begin{eqnarray}
\dfrac{2}{M}  \int_{q_\perp\,q_\parallel\,r_\perp\,r_\parallel} e^{i (q+r)x} \, \tilde {\cal G}_\Delta(q_\perp,q_\parallel)
\, \tilde S^u(r_\perp,r_\parallel) \int d^4z \,  e^{i \Phi_p(x,z)} \, e^{-i (q+r)z} \, \psi_p(z) &=&
\psi_p(x)\ ,
\nonumber \\
\dfrac{2}{M} \int_{q_\perp\,q_\parallel\,r_\perp\,r_\parallel} e^{i (q+r)x} \, \tilde {\cal G}_\Delta(q_\perp,q_\parallel)
\, \tilde S^d(r_\perp,r_\parallel) \int d^4z \, e^{-i (q+r)z} \,  \psi_n(z) &=&
\psi_n(x)\ ,
\label{barcuatro}
\end{eqnarray}
where the Schwinger phase appearing in the equation for the proton is given
by
\begin{equation}
\Phi_p(x,x') \ = \ \Phi_\Delta(x,x') + \Phi_{u}(x,x') \ = \
\dfrac{Q_p B}{2} (x_1+x'_1) (x_2-x_2')\ ,
\label{phasep}
\end{equation}
with $Q_p=e$. As expected, in the equation for the neutron the Schwinger
phase vanishes. In order to change to a momentum basis, it is convenient to
introduce the transformations
\begin{eqnarray}
\psi_p(x) & = & \sumint_{\bar P}\ \mathbb{E}_{\bar P}^p(x) \, \psi_p(\bar
P)\ , \nonumber \\
\psi_n(x) & = & \int \frac{d^4P}{(2\pi)^4} \ e^{i P x} \, \psi_n(P)\ .
\label{trans}
\end{eqnarray}
Note that while in the case of the neutron $P$ denotes the usual
four-momentum, for the proton field we have used a shorthand notation which
resembles the one used for the diquarks, namely,
\begin{equation}
\bar P \equiv (k,P_2,P_\parallel)\ ,
\qquad\qquad \sumint_{\bar P}\ \equiv \ \dfrac{1}{2\pi}\sum_{k=0}^\infty
\int \dfrac{dP_2}{2\pi}\,\int_{P_\parallel}\ .
\label{notation3}
\end{equation}
The functions $\mathbb{E}_{\bar P}^p$ are given by
\begin{equation}
\mathbb{E}_{\bar P}^p(x) \ = \sum_{\lambda=\pm} E_{\bar P,\lambda}^p(x)  \, \Gamma_\lambda \ ,
\label{EP}
\end{equation}
where $\Gamma_+ = {\rm diag}(1,0,1,0)$, $\Gamma_+ = {\rm diag}(0,1,0,1)$, and
\begin{equation}
E_{\bar P,\lambda}^p(x) \ = \ N_{k_\lambda} \, e^{i ( P_2 x_2 + P_3 x_3 + P_4 x_4)}
\, D_{k_\lambda}\left(\sqrt{2 B_p}\,x_1-s_p\sqrt{2B_p^{-1}}\,P_2\right)\ .
\label{Fq}
\end{equation}
As in the diquark case, $D_{k_\lambda}(x)$ are cylindrical parabolic
functions. We have also defined $N_{k_\lambda}= (4\pi
B_p)^{1/4}/\sqrt{k_\lambda!}$, $k_\lambda = k-(1-\lambda s_p)/2$, $B_p = |e
B|$  and $s_p=\mathrm{sign}(e B)$.

Eqs.~(\ref{barcuatro}) can be now transformed to momentum space using
Eqs.~(\ref{trans}). One gets
\begin{eqnarray}
\sumint_{\bar P'} \, \mathbb{D}^{(p)}_{\bar P \bar P'} \,
\psi_p(\bar P') & = & 0 \ , \nonumber \\
\mathbb{D}^{(n)}_{P} \, \psi_n(P) & = & 0 \ ,
\end{eqnarray}
where
\begin{eqnarray}
\mathbb{D}^{(p)}_{\bar P \bar P'} &=& \hat \delta_{\bar P \bar P'} \, \openone \,
- \, \frac{2}{M} \int_{q_\perp\,q_\parallel\,r_\perp\,r_\parallel}\,
\sum_{\lambda,\lambda'} \, I^{\lambda,\lambda'}_{\bar P \bar P'}(q,r) \,
\tilde {\cal G}_\Delta(q_\perp,q_\parallel) \,
\Gamma_\lambda \, \tilde S^u(r_\perp,r_\parallel) \, \Gamma_{\lambda'}\ ,
\label{dp}
\\
\mathbb{D}^{(n)}_{ P} &=&
\openone\, - \, \frac{2}{M} \int_{q_\perp\,q_\parallel}\tilde {\cal G}_\Delta(q_\perp,q_\parallel)
\, \tilde S^d(P_\perp-q_\perp,P_\parallel-q_\parallel)\ ,
\end{eqnarray}
with
\begin{eqnarray}
I^{\lambda,\lambda'}_{\bar P \bar P'}(q,r) =
\int d^4x \, d^4z  \ e^{i \left[ \Phi_p(x,z) + (q+r)(x-z) \right] } \,
E_{\bar P,\lambda}^p(x)^\ast\, E_{\bar P',\lambda'}^p(z)\ .
\label{illp}
\end{eqnarray}
{}From Eq.~(\ref{dp}) it is not obvious that $\mathbb{D}^{(p)}_{\bar P \bar
P'}$ is diagonal in Ritus space. However, after a rather long calculation,
it can be shown that $\mathbb{D}^{(p)}_{\bar P \bar P'}$ is indeed
proportional to $\hat \delta_{\bar P \bar P'}$. The main steps of the
calculation are given in App.~\ref{appA}. Using the form of the quark propagator
given in Eq.~(\ref{sfp_schw}) one finally obtains
\begin{eqnarray}
\mathbb{D}^{(p)}_{\bar P \bar P'} &=& \hat \delta_{\bar P \bar P'}
\sum_{\lambda = \pm} \left[ X^{(p)}_\lambda + Y^{(p)}_\lambda \, P_\parallel \cdot \gamma_\parallel
+ Z_\lambda^{(p)} \, \gamma_2\right] \Gamma_\lambda \ ,
\nonumber \\
\mathbb{D}^{(n)}_{P} &=& \sum_{\lambda = \pm} \left[ X^{(n)}_\lambda + Y^{(n)}_\lambda
\, P_\parallel \cdot \gamma_\parallel
+ Z^{(n)} \, P_\perp \cdot \gamma_\perp\right] \Gamma_\lambda\ ,
\label{barcinco}
\end{eqnarray}
where
\begin{eqnarray}
X^{(p)}_\lambda &=& 1 - \frac{8 \pi}{B_p }
 (-1)^{k_\lambda}
\int_{q_\perp\,q_\parallel\,r_\perp} e^{-(q_\perp+r_\perp)^2/B_p}\; \tilde {\cal
G}_\Delta(q_\perp,q_\parallel)\nonumber \\
& & \times \ T^u_\lambda(r_\perp,P_\parallel-q_\parallel)\,
L_{k_\lambda}\Big( \frac{2(r_\perp+q_\perp)^2}{B_p}\Big)\ ,
\label{upa} \\
Y^{(p)}_\lambda &=& \dfrac{8 \pi }{M B_p }  (-1)^{k_\lambda}
\int_{q_\perp\,q_\parallel\,r_\perp} e^{-(q_\perp+r_\perp)^2/B_p} \;
\tilde {\cal G}_\Delta(q_\perp,q_\parallel)\nonumber \\
& & \times\; \Big( 1- \dfrac{ q_\parallel \cdot P_\parallel}{P^2_\parallel} \Big)
\; T^u_\lambda(r_\perp,P_\parallel-q_\parallel)\;
L_{k_\lambda}\Big( \frac{2(r_\perp+q_\perp)^2}{B_p}\Big)\ ,
\label{upb} \\
Z_\lambda^{(p)} &=&  \dfrac{8 \pi \, s_p}{M B_p} \sqrt{ \dfrac{2}{k B_p} } \,(-1)^k
\int_{q_\perp\,q_\parallel\,r_\perp} e^{-(q_\perp+r_\perp)^2/B_p} \;
\tilde {\cal G}_\Delta(q_\perp,q_\parallel)\nonumber \\
& & \times \;
r_\perp \big[ (r_1+q_1) - i \lambda  (r_2+q_2) \big]\, V^u(r_\perp,P_\parallel-q_\parallel)
\;L^1_{k-1}\Big( \frac{2(r_\perp+q_\perp)^2}{B_p}\Big)\ ,
\label{upc} \end{eqnarray}
and
\begin{eqnarray}
X^{(n)}_\lambda &=& 1- 2
\int_{q_\perp\,q_\parallel} \ \tilde {\cal G}_\Delta(q_\perp,q_\parallel)\
T^d_\lambda(P_\perp-q_\perp,P_\parallel-q_\parallel)\ ,
\label{dpa}\\
Y^{(n)}_\lambda &=& \frac{2}{M}
\int_{q_\perp\,q_\parallel} \ \tilde {\cal G}_\Delta(q_\perp,q_\parallel)\
T^d_\lambda(P_\perp-q_\perp,P_\parallel-q_\parallel)
\Big( 1- \dfrac{ q_\parallel \cdot P_\parallel}{P^2_\parallel} \Big)\ ,
\label{dpb}\\
Z_\lambda^{(n)} &=& \frac{2}{M}
\int_{q_\perp\,q_\parallel} \ \tilde {\cal G}_\Delta(q_\perp,q_\parallel)\
V^d(P_\perp-q_\perp,P_\parallel-q_\parallel)
\Big( 1- \dfrac{ q_\perp \cdot P_\perp}{P^2_\perp} \Big)\ ,
\label{dpc}
\end{eqnarray}
with
\begin{eqnarray}
T^f_\lambda(r_\perp,r_\parallel) &=& \int_0^\infty d\tau \ e^{-\tau \phi_f(\tau,r)}\,
\Big[1+\lambda\, s_f \tanh(\tau B_f) \Big]\ ,
\nonumber \\
V^f(r_\perp,r_\parallel) &=& \int_0^\infty d\tau \ e^{-\tau \phi_f(\tau,r)}\,
\sech^2(\tau B_f)\ .
\end{eqnarray}

In what follows we will concentrate on the determination of the proton and
neutron lowest possible energies. Since these quantities are usually
interpreted as the nucleon masses, we denote them as ${\cal M}_N$, with
$N=p,n$. For the neutron we just take, as usual, $\vec P_\perp = 0$,
$P_3 = 0$, $P_4^2=-{\cal M}_n^2$. In the case of the proton, as done for the
diquarks, we consider the squared canonical momentum, $\Pi^2 = 2kB_p +
P_\parallel^2$. The lowest energy state corresponds to the lowest Landau
level (LLL), $k=0$. Then, taking $P_3 = 0$, one has $P_4^2 = - {\cal M}_p^2$, as
for the neutron case. Since the determinants of the Dirac operators in
Eqs.~(\ref{barcinco}) have to vanish at the pole masses, the
corresponding eigenvalue equations read
\begin{eqnarray}
{{}{\hat X^{(p)\,}_{s_p}}}^2 - {\cal M}^2_p\; {{}{\hat Y^{(p)\,}_{s_p}}}^2 & = & 0 \ ,
\label{eigenp} \\
{{}{\hat X^{(n)\,}_{\lambda}}}^2 - {\cal M}^2_n\; {{}{\hat Y^{(n)\,}_{\lambda}}}^2 & = &
0\ ,
\label{eigenn}
\end{eqnarray}
where we have denoted by $\hat X^{(N)}_\pm$ and $\hat Y^{(N)}_\pm$ the
coefficients in Eqs.~(\ref{barcinco}) evaluated at $k=0$, $P_3 =0$, $\vec
P_\perp =0$. Note that for the lowest energy states there is no contribution
from the terms with $Z_\lambda^{(p)}$ and $Z_\lambda^{(n)}$. In addition, in
the case of the proton only the projection $\lambda= s_p$ is nonvanishing
for $k=0$. For the neutron, both projections are in principle allowed, and
one should take the value of $\lambda$ that leads to the lowest value of the
mass.

To obtain the explicit form of the coefficients $\hat X^{(N)}_\lambda$ and
$\hat Y^{(N)}_\lambda$ needed to evaluate ---and solve---
Eqs.~(\ref{eigenp}-\ref{eigenn}) one has to replace the diquark propagator,
Eq.~(\ref{diprop}), in Eqs.~(\ref{upa}), (\ref{upb}), (\ref{dpa}) and
(\ref{dpb}). For convenience we consider first the form of the coefficients
in the absence of the external magnetic field (in this case both proton and
neutron are taken at rest). They are given by (see App.~\ref{appB})
\begin{eqnarray}
\hat X &=& 1 - \frac{1}{4 \pi^2 m_N} \int_1^\infty
\frac{d\tau}{\tau} \int_0^\infty dq\, q^2 \ {\cal G}^{\rm (reg)}_{\Delta,B=0}(q^2)\,
e^{ - \tau  (M^2 + q^2 - m_N^2) /\Lambda_B^2}\,
J_1\Big(\frac{2\, \tau\, q\, m_N}{\Lambda_B^2}\Big)\ ,
\label{AA} \\
& & \nonumber \\
\hat Y &=& \frac{1}{4 \pi^2 m_N  M} \int_1^\infty
\frac{d\tau}{\tau} \int_0^\infty dq\, q^2 \ {\cal G}^{\rm (reg)}_{\Delta,B=0}(q^2)\,
e^{ - \tau  (M^2 + q^2 - m_N^2)/\Lambda_B^2}
\nonumber \\
&& \times \left[
J_1\Big(\frac{2\, \tau\, q\, m_N}{\Lambda_B^2}\Big) - \frac{q}{m_N}
J_2\Big(\frac{2\, \tau\, q\, m_N}{\Lambda_B^2}\Big)\right]\ .
\label{BB}
\end{eqnarray}
Here, and below, $m_N$ denotes the nucleon mass at $B=0$, and $J_k(x)$ are
Bessel functions. The $B=0$ diquark propagator [see Eq.~(\ref{glq})] is
given by
\begin{equation}
{\cal G}^{\rm (reg)}_{\Delta,B=0}(q^2) \ = \
\left[ \frac{1}{4H}\, - J_{\Delta,B=0}^{{\rm (reg)}}(q^2) \right]^{-1}\ .
\end{equation}
Notice that Eqs.~(\ref{AA}) and (\ref{BB}) include a cutoff parameter
$\Lambda_B$, which has been introduced in order to regularize the otherwise
divergent quark-diquark loop within the proper time regularization scheme.

For nonzero magnetic field $B$, in the case of the proton we have
\begin{eqnarray}
\hspace{-0.4cm} \hat X^{(p)}_{s_p} &=& 1 - \frac{B_u B_\Delta}{2 \pi^2
\Lambda^2_B} \int_1^\infty d\tau \ \frac{1 + t_u}{ B_u + (B_p + B_\Delta)
t_u}\sum_{\ell = 0}^\infty
\left[\frac{ B_u + (B_p - B_\Delta) t_u}{ B_u + (B_p + B_\Delta) t_u} \right]^\ell \nonumber \\
& & \hspace{-1.2cm} \times \int_0^\infty \! dq_\parallel\, \, q_\parallel
\; {\cal G}^{\rm (reg)}_\Delta(\ell,q_\parallel^2)\;  e^{ -  \tau  (M^2 +
q_\parallel^2 - {\cal M}_p^2) /\Lambda_B^2} J_0
\Big(\frac{2\, \tau\, q_\parallel\, {\cal M}_p}{\Lambda_B^2}\Big) \ ,
\label{X1} \\ \nonumber &&
\\
\hspace{-0.4cm} \hat Y^{(p)}_{s_p} &=&  \frac{B_u B_\Delta}{2 \pi^2 M \Lambda^2_B} \int_1^\infty d\tau
\ \frac{1 + t_u}{ B_u + (B_p + B_\Delta) t_u}\sum_{\ell = 0}^\infty
\left[\frac{ B_u + (B_p - B_\Delta) t_u}{ B_u + (B_p + B_\Delta) t_u} \right]^\ell \nonumber \\
&& \hspace{-1.2cm} \times
\int_0^\infty \! dq_\parallel\, \, q_\parallel \;
{\cal G}^{\rm (reg)}_\Delta(\ell,q_\parallel^2)\ e^{ - \tau  (M^2 + q_\parallel^2 -
{\cal M}_p^2) /\Lambda_B^2} \left[
J_0\Big(\frac{2\, \tau\, q_\parallel\, m_p}{\Lambda_B^2}\Big)-
\frac{q_\parallel}{{\cal M}_p} J_1\Big(\frac{2\, \tau\, q_\parallel\,
{\cal M}_p}{\Lambda_B^2}\Big)
\right] ,
\label{fullpr}
\end{eqnarray}
while for the neutron we get
\begin{eqnarray}
\hspace{-0.4cm} \hat X^{(n)}_{\lambda} &=& 1 - \frac{B_d B_\Delta}{2 \pi^2 \Lambda^2_B} \int_1^\infty d\tau \
\frac{1 +\lambda s_d t_d}{ B_d + B_\Delta t_d}\sum_{\ell = 0}^\infty
\left[\frac{ B_d - B_\Delta t_d}{ B_d + B_\Delta t_d} \right]^\ell \nonumber \\
&&\hspace{-1.2cm} \times \int_0^\infty \! dq_\parallel\, \, q_\parallel
\; {\cal G}^{\rm (reg)}_\Delta(\ell,q_\parallel^2) \ e^{ -  \tau  (M^2 +
q_\parallel^2 - {\cal M}_n^2) /\Lambda_B^2}
J_0\Big(\frac{2\, \tau\, q_\parallel\, {\cal M}_n}{\Lambda_B^2}\Big)\ ,
\\ \nonumber &&
\\
\hspace{-0.4cm} \hat Y^{(n)}_{\lambda} &=&  \frac{B_d
B_\Delta}{2 \pi^2 M \Lambda^2_B} \int_1^\infty d\tau \ \frac{1 +
\lambda s_d t_d}{ B_d  B_\Delta t_d}\sum_{\ell = 0}^\infty
\left[\frac{ B_d - B_\Delta t_d}{ B_d + B_\Delta t_d} \right]^\ell \nonumber \\
& & \hspace{-1.2cm} \times
\int_0^\infty \! dq_\parallel\, \, q_\parallel \, {\cal
G}^{\rm (reg)}_\Delta(\ell,q_\parallel^2) \ e^{ -  \tau  (M^2 + q_\parallel^2 -
{\cal M}_n^2) /\Lambda_B^2}
\left[ J_0\Big(\frac{2\, \tau\, q_\parallel\, {\cal M}_n}{\Lambda_B^2}\Big)- \frac{q_\parallel}{{\cal M}_n}
J_1\Big(\frac{2\, \tau\,  q_\parallel\, {\cal M}_n}{\Lambda_B^2}\Big) \right] .
\label{fullne}
\end{eqnarray}
In these equations we have used the definition $t_f=\tanh (\tau
B_f/\Lambda_B^2)$.

\subsection{Nucleon magnetic moments}

We finish this section by noting that given the above expressions for
$\hat X^{(N)}_\lambda$ and $\hat Y^{(N)}_\lambda$, they can be
expanded around $B=0$ in order to study how nucleon masses get modified to
lowest order in the magnetic field. Let us define the corresponding slopes
$\alpha_N$ by
\begin{equation}
{\cal M}_N = m_N + \alpha_N\, |B| + {\cal O}(B^2)\ .
\end{equation}
After a rather long calculation, sketched in App.~\ref{appB}, we obtain
\begin{eqnarray}
\alpha_p & = & \dfrac{ - Q_u \big[ (M+m_N)\, {\cal I}_1
- m_N\,{\cal I}_2 \big] + Q_p\, \hat W}{ M \,\hat Y + 2\, m_N \, \hat W}\ , \nonumber \\
\alpha_n & = & \dfrac{ Q_d \big[ (M+m_N)\, {\cal I}_1
- m_N\,{\cal I}_2 \big]}{ M \,\hat Y + 2\, m_N \, \hat W}\ ,
\label{alphan}
\end{eqnarray}
where we have defined
\begin{eqnarray}
\hat W \ = \ (M+m_N)\, {\cal I}_1 - ( 2\, m_N + M )\, {\cal I}_2 + m_N \, {\cal
I}_3\ ,
\end{eqnarray}
and the integrals ${\cal I}_k$ are given by
\begin{eqnarray}
{\cal I}_k = \frac{1}{4 \pi^2\,\Lambda_B^2\,m_N^k} \int_1^\infty
d\tau \int_0^\infty dq\, q^{k+1} \, {\cal G}_{\Delta,B=0}(q^2)\,
e^{ - \tau  (M^2 + q^2 - m_N^2)/\Lambda_B^2}\,
J_k\Big(\frac{2\, \tau\,  q\, m_N}{\Lambda_B^2}\Big)\ .
\label{ik}
\end{eqnarray}

To find the relation between $\alpha_N$ and the nucleon magnetic
moments we proceed as follows. First, we take into account that to leading
order in the magnetic field the change in the nucleon energy is given by
\cite{Tiburzi:2008ma,Primer:2013pva}
\begin{equation}
\Delta E_N \ = \ \frac{ |Q_NB|}{2 m_N} - \vec \mu_N \cdot \vec B + {\cal
O}(B^2) \ .
\label{unos}
\end{equation}
The first term corresponds to orbital motion. While it vanishes for the neutron,
for the proton it provides a
contribution due to zero point motion in the plane perpendicular to the
magnetic field. The second term represents, for both $p$ and $n$, the spin
contribution leading to the Zeeman effect. Thus, we have
\begin{eqnarray}
\Delta E_p &=& (1 -  \mu_p) \ \frac{e|B|}{2 m_N}+ {\cal O}(B^2) \ ,
\nonumber \\
\Delta E_n &=& - \lambda \ \mu_n \ \frac{eB}{2 m_N}+ {\cal O}(B^2) \ ,
\end{eqnarray}
where, as usual, the nucleon magnetic moments are expressed in units of the
nuclear magneton $\mu_N = e/(2 m_N)$. Note that for the proton we have taken
into account the fact that for the lowest energy state one has
$\lambda=s_p$. In this way, identifying the corresponding slopes at $B=0$,
the nucleon magnetic moments are given by
\begin{eqnarray}
\mu_{p} &=&  1 - \frac{2 m_N}{e}\; \alpha_p\ , \nonumber \\
\mu_{n} &=& - \lambda\, {\rm sign}(B)\, \frac{2 m_N}{e} \; \alpha_n\ .
\label{magmomne}
\end{eqnarray}

\section{Numerical results}

\begin{figure}[h]
    \centering{}\includegraphics[width=0.59\textwidth]{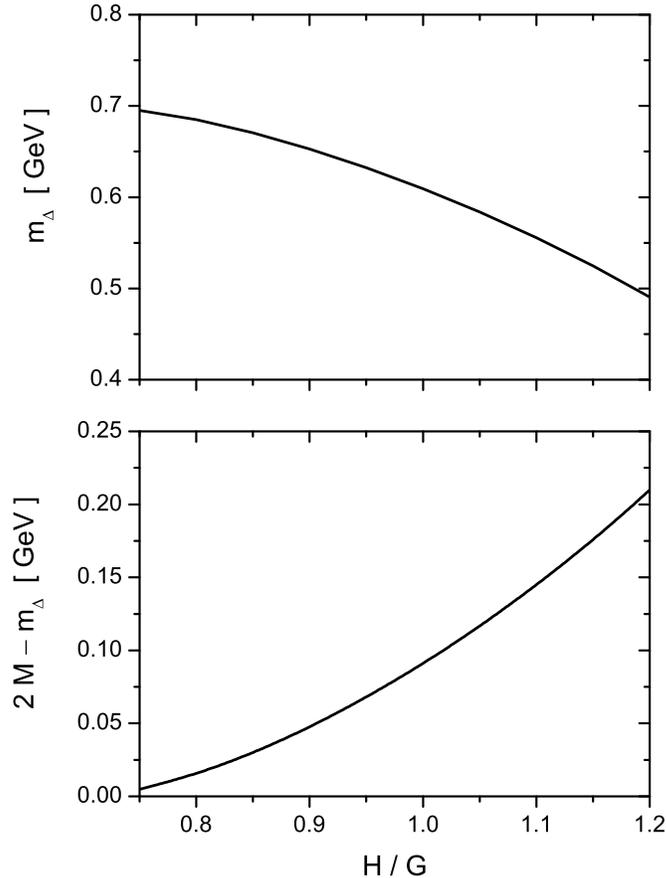}
\caption{$\Delta$ mass (top) and binding energy (bottom) at $B=0$ as functions of $H/G$.}
\label{fig:2}
\end{figure}

To obtain numerical results for diquark and baryon properties one has to fix
the model parametrization. Here, as done in Ref.~\cite{Coppola:2018vkw}, we
take the parameter set $m_0 = 5.66$~MeV, $\Lambda = 613.4$~MeV and
$G\Lambda^2 = 2.250$, which (for vanishing external field) corresponds to a
constituent quark mass $M=350$~MeV and a quark-antiquark condensate $\langle
\bar f f\rangle = (-243.3\ {\rm MeV})^3$. This parametrization properly
reproduces the empirical values of the pion mass and decay constant in
vacuum, $m_\pi=138$~MeV and $f_\pi=92.4$~MeV. It also provides a very good
agreement with the results from lattice QCD quoted in
Ref.~\cite{Bali:2011qj} for the normalized average $\bar ff$ condensate,
$\Delta \bar\Sigma(B)$, up to $|eB|\simeq 1$~GeV$^2$~\cite{Coppola:2018vkw}.
The effective Lagrangian in Eq.~(\ref{lagrangian}) also includes the scalar
quark-quark coupling constant $H$. Typical effective approaches for the
strong interaction, such as the One Gluon Exchange or the Instanton Liquid
Model, lead to $H/G = 0.75$~\cite{Buballa:2003qv}. However, this value
is subject to somewhat large uncertainties from the phenomenological point
of view. In fact, larger values for this ratio seem to be favored from the
determination of baryon properties within the Fadeev
approach~\cite{Buck:1992wz,Huang:1993yd,Ishii:1993np,Ishii:1995bu}. Here we
choose to take $H/G$ within the range $0.75 \leq H/G \leq 1.2$, typically
considered in the literature. The corresponding values of the diquark mass
and binding energies are shown in Fig.~\ref{fig:2}. We observe that for $H/G
\simeq 0.75$ the scalar diquark is barely bound by 5~MeV, while for $H/G =
1.2$ one gets binding energies of about 200 MeV.

\begin{figure}[h]
    \centering{}\includegraphics[width=0.59\textwidth]{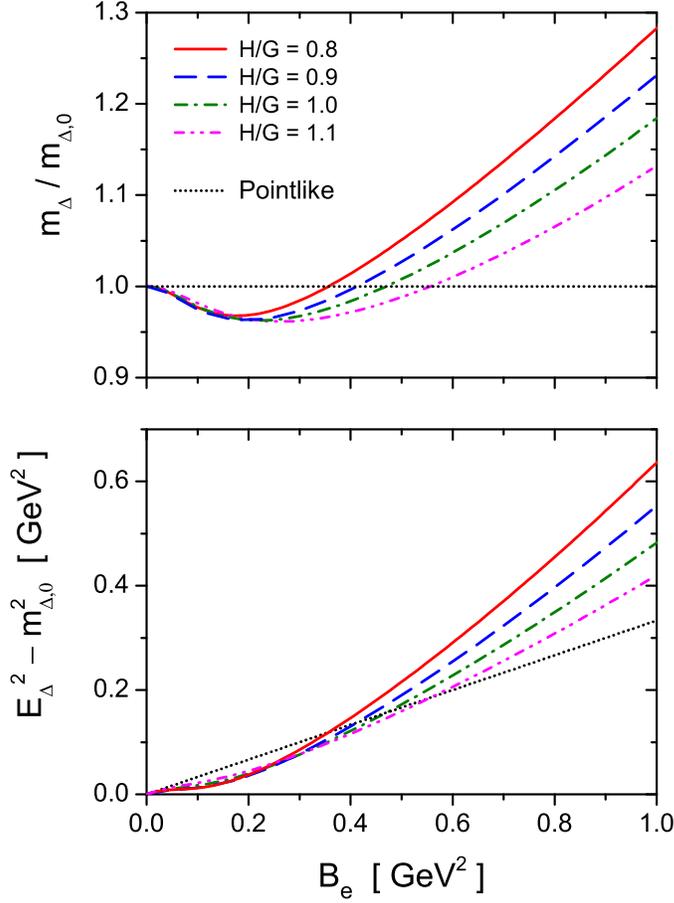} \caption{(Color
online) Relative values of diquark mass and energy as functions of $|eB|$
for some representative values of $H/G$. The results for the case of a
pointlike diquark are indicated by the dotted lines.} \label{fig:02}
\end{figure}

Let us consider the magnetic field dependence of the diquark mass. In the
upper panel of Fig.~\ref{fig:02} we show the values of $m_\Delta$ for the
LLL [defined by Eq.~(\ref{polediquark}), with $\ell=0$] relative to the
values obtained for vanishing magnetic field, $m_{\Delta,0}$, as functions
of $B_e = |eB|$. The curves correspond to some selected values of the ratio
$H/G$ within the range mentioned above. We get $m_{\Delta,0} = 0.685$,
$0.653$, $0.609$ and $0.555$~GeV for $H/G=0.8$, $0.9$, $1.0$ and $1.1$,
respectively. It is seen that for all considered values of $H/G$ the curves
start with a decrease of $m_\Delta$ as $B_e$ increases, reaching a minimum
at about $B_e \sim 0.2$~GeV$^2$. Beyond this minimum the diquark pole mass
steadily increases with the magnetic field, reaching a ratio
$m_\Delta/m_{\Delta,0} = 1$ somewhere in the range $B_e \sim
0.4-0.6$~GeV$^2$, depending on the precise value of $H/G$. In the lower
panel of Fig.~\ref{fig:02} we show the behavior of the squared ``magnetic
field-dependent'' diquark mass, $E_\Delta^2$ [defined by
Eq.~(\ref{epimas})], minus the corresponding value at $B=0$,
$m_{\Delta,0}^2$. We recall that in the case of a pointlike diquark the mass
$m_{\Delta}$ does not depend on the magnetic field, and the difference
$E_\Delta^2 - m_{\Delta,0}^2$ is simply given by $B_e/3$. Such a case is
indicated by the straight dotted black line. It can be observed that, as a
consequence of the initial decrease of the pole mass, for small values of
$B_e$ the difference $E_\Delta^2 -  m^2_{\Delta,0}$ lies below that straight
line. At the point in which $m_\Delta = m_{\Delta,0}$ the situation
reverses, and for larger values of $B_e$ the value of $E_\Delta$ becomes
larger than in the case of a pointlike diquark. We notice that a similar
behavior was found in the analysis of Ref.~\cite{Liu:2018zag}, where
Schwinger phases were not taken into account. However, in that work the
crossing was found to occur at a larger value of $B_e$, of about 0.9~GeV$^2$
for $H/G=0.75$. It is interesting to note that as $H/G$ increases the
behavior of $E_\Delta^2 - m^2_{\Delta,0}$ gets closer to the pointlike case.
This might be understood by realizing that a larger value of $H/G$ implies a
more deeply bound diquark and, consequently, a more localized one.

We turn next to the analysis of nucleon masses. As mentioned in
Sec.~III.C, the calculation of these quantities requires the introduction of
an additional cutoff parameter, $\Lambda_B$, to regularize the otherwise
divergent quark-diquark loop in the
propertime regularization scheme. For a given value of $H/G$, we adjust this
parameter demanding the $B=0$ eigenvalue equation $|\hat X| = m_N |\hat Y|$
[see Eqs.~(\ref{AA}) and (\ref{BB})] to be satisfied for the physical value
$m_N = 0.938$~GeV. In this way we obtain $\Lambda_B=1.618$, 1.380 and
1.104~GeV for $H/G=0.8$, 0.9 and 1.0, respectively. For larger values of
$H/G$, no value of $\Lambda_B$ is found to be compatible with the physical
nucleon mass at zero magnetic field in this model. Having determined all
input parameters, one can solve the eingenvalue equations (\ref{eigenp}-\ref{eigenn}) to
obtain proton and neutron masses for nonvanishing external magnetic field.

Before reporting the corresponding results, we find it convenient to make a
few comments concerning the numerical details of the calculation. Firstly,
we note that to evaluate the coefficients $\hat X_\pm^{(N)}$ and $\hat
Y_\pm^{(N)}$ in Eqs.~(\ref{X1}-\ref{fullne}) one has to perform a sum over
Landau levels (LL). In that sum we have taken into account as many LL as
needed in order to obtain a stable result for the calculated mass. For low
values of $B_e$, this implies the inclusion of a quite large number of LL.
For example, at $B_e=0.04$~GeV$^2$, for $H/G=1$ about 300 LL are needed in
order to obtain an accuracy of about 1~MeV in the nucleon mass. For
$H/G=0.8$ the required number of LL is found to be even larger, of the order
of 600. As expected, for larger values of the magnetic field the needed
number of LL gets significantly reduced. Still, it is found that
for $B_e$ as large as 0.8~GeV$^2$ about 10 LL are needed to obtain
the above mentioned accuracy in the mass determination. Another issue that
requires some care is the numerical evaluation of the integrals in
Eqs.~(\ref{X1}-\ref{fullne}), due to the highly oscillatory behavior of the
Bessel functions for large values of their arguments.

\begin{figure}[ht]
\centering{}\includegraphics[width=0.7\textwidth]{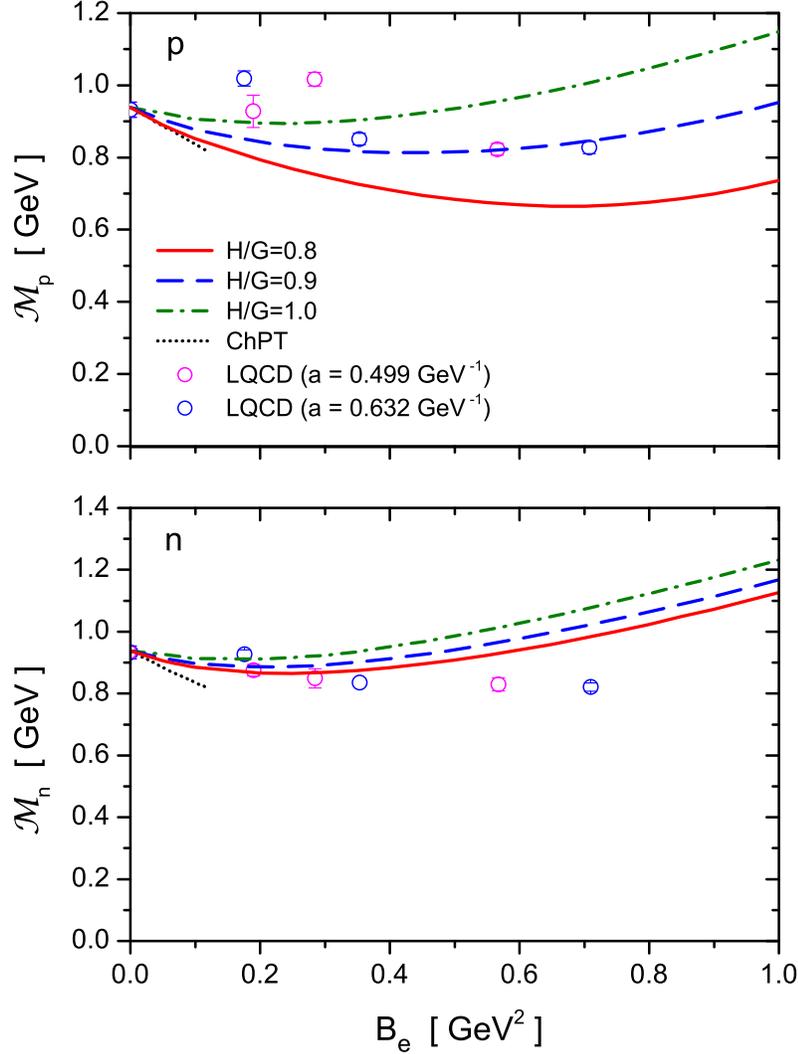}
\caption{(Color online) Proton and neutron masses as functions of
$|eB|$ for various values of $H/G$. Open dots and dotted lines correspond to
Lattice QCD results given in Ref.~\cite{Endrodi:2019whh} and ChPT results
given in Ref.~\cite{Deshmukh:2017ciw}, respectively.}
\label{fig:03}
\end{figure}

Our results for the behavior of nucleon masses as functions of the external
magnetic field are given in Fig.~\ref{fig:03}. In the upper (lower) panel we
quote the curves for the proton (neutron) mass, considering $H/G = 0.8$, 0.9
and 1.0. In all cases it is seen that the masses initially decrease when the
magnetic field is increased, reaching a minimum for a value of $B_e$ that
depends on the parameter $H/G$. Beyond that point, the masses show a steady
growth. For both proton and neutron masses, the decrease becomes less
pronounced (and the minimum occurs at smaller $B_e$) the larger the value of
$H/G$ is. It is also seen that the dependence on $H/G$ is weaker in the case
of the neutron. Let us recall that for a proton in the LLL only the spin
projection $\lambda=s_p = \mbox{sign}(Q_p B)$ is allowed, while both values
of $\lambda$ are allowed for the neutron. In Fig.~\ref{fig:03} we have
plotted the values corresponding to the lower solution of
Eq.~(\ref{eigenn}), defined as the neutron mass. In our model, for $B > 0\
(B < 0)$ it is found that this lower state corresponds to $\lambda=-1\
(\lambda=1)$. For the higher state, not shown in the figure, it is seen that
the value of ${\cal M}_n$ obtained as a solution of Eq.~(\ref{eigenn})
initially increases with $B_e$. This solution is found to exist only for
$B_e \lesssim 0.1-0.2$~GeV$^2$ (the state becomes unbound for
larger values of the external field).

As stated, close to $B=0$ both proton and neutron masses are shown to
decrease for increasing external field, i.e.\ the slopes $\alpha_p$ and
$\alpha_n$ obtained from Eq.~(\ref{alphan}) are found to be negative. Taking
into account that for the lowest neutron state one has $\lambda\, {\rm
sign}(B) = -1$, from Eqs.~(\ref{magmomne}) one gets $\mu_p > 0$ and $\mu_n <
0$, as expected from phenomenology. In addition, the fact that the curves
show negative slopes at $B = 0$ is consistent with the results from ChPT
quoted in Ref.~\cite{Deshmukh:2017ciw}. The latter, which are expected to
hold for low values of the external field, are shown by dotted lines in
Fig.~\ref{fig:03}. Notice, however, that the slopes obtained within ChPT are
in general steeper that those found from our results. The lower slopes in
our model imply in turn relatively low results for the absolute values of
proton and neutron magnetic moments. From the numerical evaluation of
Eqs.~(\ref{alphan}) and (\ref{magmomne}) we find the magnetic moments quoted
in Table~\ref{Tmag}, to be compared with the empirical values $\mu_p = 2.79$
and $\mu_n = -1.91$. In this regard, it should be stressed that in our work
we have neglected for simplicity the axial vector diquark correlations. The
latter can be important to get an enhancement in $|\mu_p|$ and $|\mu_n|$, as
shown in Ref.~\cite{Mineo:2002bg}. Finally, let us compare our results with
those obtained from LQCD calculations. In Fig.~\ref{fig:03} we have
indicated with open dots the results from LQCD quoted in
Ref.~\cite{Endrodi:2019whh}, corresponding to two different values of the
lattice spacing $a$. We observe some qualitative agreement with our results,
although LQCD values tend to show a lower dependence on the external field.
In the case of the proton, a few lattice points seem to show a mass
enhancement for $B_e\simeq 0.2 - 0.3$~GeV$^2$. Presumably, this could be due
to the fact that, as mentioned by the authors of
Ref.~\cite{Endrodi:2019whh}, the Zeeman-splitting cannot be fully resolved.
We believe that our results exhibit a more trustable initial slope, in view
of the results arising from ChPT.

\begin{table}[h]
\begin{center}
\setlength{\tabcolsep}{20pt}
\begin{tabular}{ c | c | c }
\hline \rule{0pt}{.9\normalbaselineskip}
$H/G$ & $\mu_p$ & $\mu_n$  \\ [.2cm]
 \hline  \hline \rule{0pt}{.9\normalbaselineskip}
 0.8 & 2.63 & -1.19  \\[.2cm]
 \hline \rule{0pt}{.9\normalbaselineskip}
 0.9 & 2.30 & -1.05 \\[.2cm]
 \hline \rule{0pt}{.9\normalbaselineskip}
 1.0 & 1.99 & -0.94 \\[.2cm]
 \hline
\end{tabular}

\hfill

\caption{Predicted values of nucleon magnetic moments for different values of $H/G$.}
\label{Tmag}
\end{center}
\end{table}

\section{Summary and Conclusions}

In this work we have explored the effect of a strong external uniform
magnetic field on diquark and nucleon masses. This has been done in the
framework of a two flavor Nambu-Jona--Lasinio effective model for low energy
QCD dynamics, including scalar quark-quark color pairing interactions to
account for the diquarks. The relative strength of these interactions is
determined by a coupling constant ratio $H/G$, where $H$ and $G$ are
the coupling constants driving the scalar quark-quark and pseudoscalar 
quark-antiquark interactions, respectively. We have considered values of this
ratio in the usually studied range $0.75\le H/G \le 1.2$.

As done in the case of pions, diquarks have been treated as quantum
fluctuations in the random phase approximation. Due to the presence of the
external field, translational invariance turns out to be broken, as signaled
by the presence of non-vanishing Schwinger phases, and the usual momentum
basis cannot be used to diagonalize the corresponding polarization function.
A proper basis can be found following the method introduced in
Ref.~\cite{Coppola:2018vkw} for charged pions, based on the Ritus
eigenfunction approach to magnetized relativistic systems. In view of the
non-renormalizability of the NJL model, we have adopted as regularization
procedure the Magnetic Field Independent Regularization scheme, as suggested
from the scheme comparison performed in Ref.~\cite{Avancini:2019wed}. From
the regularized diagonal polarization function we have obtained the lowest
Landau level diquark pole mass $m_\Delta$ and the ``magnetic field-dependent
mass'' $E_\Delta$, defined as the lowest quantum-mechanically allowed
diquark energy. The numerical results for these quantities show that for low
values of $|eB|$ the curves for both $m_\Delta$ and $E_\Delta$ lie below
those corresponding to a pointlike diquark. This is reversed for $|eB|$
larger than $\sim 0.3 - 0.5$~GeV$^2$, where the growth of $E_\Delta$ gets
steeper in comparison with the pointlike case. It is also found that the
increase of the ``magnetic field-dependent mass'' becomes more pronounced
for lower values of the ratio $H/G$.

Regarding the analysis of baryon states, in our framework nucleons have
been built as bound quark-diquark states following a relativistic Fadeev
approach in which only the formerly discussed scalar diquark channel is
included. Given the complexity of the problem, we have considered a static
approximation in which one disregards the momentum dependence of the exchanged quark. This
approximation has been shown to lead to an adequate description of nucleon
properties in the absence of external fields~\cite{Ishii:1993np}. Once
again, owing to the presence of nonvanishing Schwinger phases for charged particles, in the
theoretical analysis we have made use of the Ritus eigenfunction method.
In addition, we have introduced a further model parameter $\Lambda_B$
to regularize the otherwise divergent quark-diquark loops, for which we have
chosen the proper time regularization scheme. We have found that for
values of $H/G$ larger than 1 no value of $\Lambda_B$ is compatible with a
physical value of the nucleon mass at zero external magnetic field.

We have obtained numerical results for the magnetic field dependence of the
lowest energy nucleon states, usually interpreted as the nucleon masses. In
general, it is seen that the masses initially decrease for increasing
magnetic field, whereas they show a steady growth for large values of
$|eB|$. In the case of the proton the results are found to depend strongly
on the ratio $H/G$. It is also seen that the negative slopes of the mass
curves at $B=0$ lead to the phenomenologically correct signs for the nucleon
magnetic moments. Moreover, there is a qualitative agreement with ChPT
results, although the slopes in our model are found to be somewhat lower.
This conduces to numerical absolute values for the proton and neutron
magnetic moments that are relatively small in comparison with the empirical
ones.

The work presented in this article represents a first approach to
relativistic magnetized nucleons as bound quark-diquark states within the
NJL model. An improvement on the predictions for the nucleon magnetic
moments is expected to be obtained by including axial vector diquark
interactions. Moreover, a full calculation would require to take into
account the momentum dependence of the exchanged quark. We expect to report
on these issues in future publications.

\acknowledgments

We thank S.~Noguera for useful discussions. This work has been supported in
part by Consejo Nacional de Investigaciones Cient\'ificas y T\'ecnicas and
Agencia Nacional de Promoci\'on Cient\'ifica y Tecnol\'ogica (Argentina),
under Grants No.~PIP17-700 and No.~PICT17-03-0571, respectively, and by the
National University of La Plata (Argentina), Project No.~X824.

\appendix

\section{Diagonalization of $\mathbb{D}^{(p)}_{\bar P \bar P'}$ in Ritus space}
\label{appA}

In this appendix we briefly sketch how to prove that the Dirac operator
$\mathbb{D}^{(p)}_{\bar P \bar P'}$ in Eq.~(\ref{dp}) is diagonal. Let us
start by taking into account the integral $I^{\lambda,\lambda'}_{\bar P \bar
P'}(q,r)$ in Eq.~(\ref{illp}). Denoting $w = x_1-z_1$ and integrating over
the remaining space variables, it is easy to show that
\begin{equation}
I^{\lambda,\lambda'}_{\bar P \bar P'}(q,r) \ = \ (2\pi)^6 \, \delta^{(2)}(P_\parallel-P_\parallel') \,
\delta(P_2-P_2')\, \delta^{(2)}(q_\parallel+r_\parallel-P_\parallel)\;
G_{k_\lambda,k'_{\lambda'}}(q_\perp + r_\perp)\ ,
\label{I6}
\end{equation}
where
\begin{align}
G_{k_\lambda,k'_{\lambda'}}(q_\perp + r_\perp) \ = \ &  \dfrac{(-1)^{k_{\lambda}+k'_{\lambda'}}}{B_p}
\int_0^\infty dw\; e^{i(q_1+r_1)w} \, N_{k_\lambda} \, D_{k_\lambda}\left(s_p\sqrt{2/B_p}\,(q_2+r_2)-\sqrt{B_p/2}\,w \right) \nonumber\\
& \times   N_{k'_{\lambda'}} \, D_{k'_{\lambda'}}\left( s_p\sqrt{2/B_p}\,(q_2+r_2) + \sqrt{B_p/2}\,w
\right)\ .
\end{align}
The integral over $w$ can be carried out using the following property,
\begin{eqnarray}
\hspace{-1cm}\int_0^\infty \! d\psi \, e^{i\gamma\psi} D_\ell(\eta-\psi) \, D_{n}(\eta+\psi) & = & \nonumber\\
&& \hspace{-4cm}\begin{cases}
(-1)^\ell \, \sqrt{2\pi} \, \ell! \, e^{-\frac{\gamma^2+\eta^2}{2}}\left(i\gamma+\eta\right)^{n-\ell} \, L_\ell^{n-\ell}(\eta^2+\gamma^2) \qquad \qquad  & \mathrm{if} \,\, n\geq \ell \\
(-1)^{n} \, \sqrt{2\pi} \, n! \, e^{-\frac{\gamma^2+\eta^2}{2}}\left(-i\gamma+\eta\right)^{\ell-n} \, L_{n}^{\ell-n}(\eta^2+\gamma^2)
 &  \mathrm{if} \,\, \ell\geq n \ \ .
\end{cases}
\label{prop1}
\end{eqnarray}
Assuming that $k'_{\lambda'} \ge k_\lambda$ (the analysis is similar
for the other case), one has
\begin{align}
G_{k_\lambda,k'_{\lambda'}}(q_\perp + r_\perp) =&  (-)^{k'_{\lambda'}} \, \dfrac{4\pi}{B_p} \sqrt{\dfrac{k_{\lambda}!}{k'_{\lambda'}!}} \:
e^{-\frac{(q_\perp+r_\perp)^2}{B_p}} \left[ \dfrac{i(q_1+r_1)+s_p(q_2+r_2)}{\sqrt{B_p/2}} \right]^{k'_{\lambda'}-k_{\lambda}}
\nonumber \\
& \times L_{k_{\lambda}}^{k'_{\lambda'}-k_{\lambda}} \left( \dfrac{2(q_\perp+r_\perp)^2}{B_p} \right) \ .
\label{gkkp}
\end{align}

Now let us take this result to carry out the integral over perpendicular
momenta in Eq.~(\ref{dp}),
\begin{equation}
I_\perp \ = \ \int_{q_\perp\,r_\perp}\,
\sum_{\lambda,\lambda'} \, G_{k_\lambda,k'_{\lambda'}}(q_\perp + r_\perp) \,
\tilde {\cal G}_\Delta(q_\perp,q_\parallel) \,
\Gamma_\lambda \, \tilde S^u(r_\perp,P_\parallel-q_\parallel) \, \Gamma_{\lambda'}\ .
\end{equation}
Using the form of the quark propagator in Eq.~(\ref{sfp_schw}), it can be
seen that the product $\Gamma_\lambda \, \tilde S^u(r_\perp,P_\parallel-q_\parallel) \,
\Gamma_{\lambda'}$ can be written as
\begin{equation}
\Gamma_\lambda \, \tilde S^u(r_\perp,P_\parallel-q_\parallel) \,
\Gamma_{\lambda'}\ = \
{\cal A}(r_\perp,P_\parallel-q_\parallel)\,
\delta_{\lambda\lambda'}\,\Gamma_\lambda\, + \,
{\cal B}(r_\perp,P_\parallel-q_\parallel)\,r_\perp\cdot\gamma_\perp\,
\delta_{-\lambda\lambda'}\,\Gamma_{-\lambda} \ ,
\end{equation}
where ${\cal A}(r_\perp,P_\parallel-q_\parallel)$ and ${\cal
B}(r_\perp,P_\parallel-q_\parallel)$ are functions of $r_\perp^2$. Then
we get
\begin{eqnarray}
I_\perp & = & \int_{q_\perp\,r_\perp}\,\tilde {\cal G}_\Delta(q_\perp,q_\parallel) \,
\sum_{\lambda} \, \Big[ G_{k_\lambda,k'_{\lambda}}(q_\perp + r_\perp) \,
{\cal A}(r_\perp,P_\parallel-q_\parallel)\, \Gamma_\lambda\, \nonumber \\
& &  +
%\,i\,\lambda\, G_{k_\lambda,k'_{-\lambda}}(q_\perp + r_\perp)\,
%{\cal B}(r_\perp,P_\parallel-q_\parallel)\,
%(r_1-i\lambda\, r_2)\gamma_2\Gamma_{-\lambda}\Big] \ .
\, G_{k_\lambda,k'_{-\lambda}}(q_\perp + r_\perp)\,
{\cal B}(r_\perp,P_\parallel-q_\parallel)\,
(r_1-i\lambda\, r_2)\gamma_\lambda\,\Gamma_{-\lambda}\Big] \ ,
\label{iperp}
\end{eqnarray}
where $\gamma_\lambda=(\gamma_1 + i\lambda\gamma_2)/2$. To carry out the angular integrals in Eq.~(\ref{iperp}) it is convenient to
use polar coordinates, namely $\vec q_\perp = (\tilde q \cos\theta,\tilde q \sin\theta)$,
$\vec r_\perp = (\tilde r\cos\varphi,\tilde r\sin\varphi)$. Noticing that the diquark
propagator depends only on the squared momenta $q_\parallel^2$ and $q_\perp^2$ [see
Eq.~(\ref{diprop})], from Eq.~(\ref{gkkp}) we get
\begin{align}
I_\perp  = & \int_0^\infty \dfrac{\tilde q\, d\tilde q}{(2\pi)^2} \int_0^\infty \dfrac{\tilde r\, d\tilde r}{(2\pi)^2}
\  \tilde {\cal G}_\Delta(\tilde q,q_\parallel) \, \sum_{\lambda}  \nonumber \\
& \times \Big[ {\cal A}(\tilde r,P_\parallel-q_\parallel)\, \Gamma_\lambda 
\int_0^{2\pi}\!d\varphi\, e^{-is_p(k'_\lambda-k_\lambda)\varphi}\, \int_0^{2\pi}\!d\theta\,
F_{k_\lambda,k'_{\lambda}}(\tilde q,\tilde r,\theta -\varphi) \, + \nonumber \\
& \hphantom{\times \quad} \tilde r\, {\cal B}(\tilde r,P_\parallel-q_\parallel)\,\gamma_\lambda\,
\Gamma_{-\lambda} \int_0^{2\pi} \!d\varphi\, e^{-i[s_p(k'_{-\lambda}-k_\lambda)+\lambda]\varphi} %\,e^{-i\lambda\varphi}
\int_0^{2\pi}\!d\theta\, F_{k_\lambda,k'_{-\lambda}}(\tilde q,\tilde r,\theta -\varphi)\Big]\ ,
\end{align}
where $F_{k_\lambda,k'_{\lambda'}}$ is a function that depends on
$\theta-\varphi$ only through periodic functions $\sin(\theta-\varphi)$,
$\cos(\theta-\varphi)$. Taking into account that
\begin{equation}
k'_\lambda - k_\lambda = k'-k \ ,\qquad\qquad
s_p(k'_{-\lambda}-k_\lambda) + \lambda = s_p(k'-k) \ ,
\end{equation}
and using the periodicity of the function $F_{k_\lambda,k'_{\lambda'}}$, it
is seen that $I_\perp$ is proportional to
\begin{equation}
\int_0^{2\pi} d\varphi\, e^{-i s_p (k'-k)} \ = \ 2\pi\, \delta_{kk'} \ .
\end{equation}
Together with the result in Eq.~(\ref{I6}), this shows that $\mathbb{D}^{(p)}_{\bar P \bar
P'}$ is proportional to $\hat \delta_{\bar P\bar P'}$.

\section{Expansion around $B=0$}
\label{appB}

In this appendix we provide some hints for the expansions of the coefficients
$\hat X^{(N)}_\pm$ and $\hat Y^{(N)}_\pm$ in
Eqs.~(\ref{eigenp}-\ref{eigenn}) around $B=0$. These expansions allow us to
obtain the expressions for $\hat X$ and $\hat Y$ in
Eqs.~(\ref{AA}-\ref{BB}), as well as the slopes $\alpha_N$ in
Eqs.~(\ref{alphan}).

The coefficients $\hat X^{(N)}_\pm$ and $\hat Y^{(N)}_\pm$ depend on $B$
both explicitly and implicitly, through ${\cal M}_N$ and $M$. In fact, it
can be seen that $dM/dB|_{B=0} = 0$, hence the effective quark mass $M$ can
be taken as a constant at the lowest order in an expansion in powers of
$|B|$. In this way, from Eqs.~(\ref{eigenp}-\ref{eigenn}) the slopes
$d{\cal M}_N/d|B|$ at $B=0$ are given by
\begin{equation}
\alpha_N \ = \ \frac{ \frac{\partial\hat X^{(N)}_\lambda}{\partial |B|}\big|_{B=0} - m_N\;
\frac{\partial\hat Y^{(N)}_\lambda}{\partial |B|}\big|_{B=0}}
{\hat Y - \frac{\partial \hat X}{\partial m_N} +
m_N\; \frac{\partial \hat Y}{\partial m_N}}\ ,
\label{alphaapp}
\end{equation}
where appropriate values of $\lambda$ should be taken for $N=p$ and $N=n$
(see discussion in the main text).

In particular, the partial derivatives in the numerator of the rhs of
Eq.~(\ref{alphaapp}) have to be calculated with some care due to the sums
over Landau levels in Eqs.~(\ref{X1}-\ref{fullne}). As an example, let us
consider the expression for $\hat X^{(p)}_{s_p}$ in Eq.~(\ref{X1}). The
factors that depend explicitly on the magnetic field can be expanded as
\begin{eqnarray}
\frac{B_u\,(1+t_u)}{B_u + (B_p + B_\Delta)\, t_u} & = & 1 + \frac{\tau}{\Lambda_B^2}\,(B_u - B_p -
B_\Delta)\, + {\cal O}(B^2) \nonumber \\
\rule{0cm}{0.9cm} B_\Delta \left[\frac{ B_u + (B_p - B_\Delta) t_u}{ B_u + (B_p + B_\Delta) t_u}
\right]^\ell & = & B_\Delta\,e^{-2\tau\ell B_\Delta/\Lambda_B^2}\,\big[1 + \frac{2\tau^2 \ell B_\Delta B_p}{\Lambda_B^4} +
{\cal O}(B^2,\ell B^3)\big] \nonumber \\
\rule{0cm}{0.7cm} {\cal G}^{\rm (reg)}_\Delta(\ell,q_\parallel^2) & = &
{\cal G}^{\rm (reg)}_{\Delta,B=0}(q_\parallel^2+2\ell B_\Delta) \nonumber \\
& & \rule{0cm}{0.9cm} + \; \frac{d{\cal G}^{\rm (reg)}_{\Delta,B=0}(q^2)}{dq^2}\Big|_{q^2 = q_\parallel^2+2\ell B_\delta}
\, B_\Delta + {\cal O}(B^2, \ell B^3)\ .
\end{eqnarray}
For the evaluation of the sum over Landau levels in the limit of low
magnetic field, one can use the relation
\begin{equation}
B\, \sum_{\ell=0}^\infty e^{-\alpha\, \ell B}\, F(\ell B) \ = \ \int_0^\infty
dx\ e^{-\alpha\, x} \ F(x) + \frac{1}{2}\, F(0)\, B + {\cal O}(B^2)\ ,
\end{equation}
which is valid for $\alpha>0$ if the function $F(x)$ allows a Taylor
expansion around $x=0$ and is well behaved at $x\to \infty$. In this way,
after an integration by parts one arrives at
\begin{eqnarray}
\hspace{-1cm} B_\Delta\,\frac{B_u\,(1+t_u)}{B_u + (B_p + B_\Delta)}\,\sum_{\ell =
0}^\infty\,\left[\frac{ B_u + (B_p - B_\Delta) t_u}{ B_u + (B_p + B_\Delta) t_u}
\right]^\ell\,{\cal G}^{\rm (reg)}_\Delta(\ell,q_\parallel^2) & = &
\nonumber \\
& & \hspace{-9.5cm}\frac{1}{2}\int_0^\infty d\omega\, e^{-\tau\omega/\Lambda_B^2}
\, {\cal G}^{\rm (reg)}_{\Delta,B=0}(q_\parallel^2+\omega)\,\Big[ 1 + \frac{\tau}{\Lambda_B^2}\,
(B_u-B_p) + \frac{\omega\,\tau^2\,B_p}{\Lambda_B^4} + {\cal O}(B^2)\,\Big]\ .
\label{landausum}
\end{eqnarray}

The variable $\omega$ can be identified with the perpendicular component of
the momentum squared, $q_\perp^2$, in the $B\to 0$ limit. In addition, with the aid of
some properties of the Bessel functions one can prove the relations
\begin{eqnarray}
\int_0^\infty dq_\parallel^2\ \int_0^\infty dq_\perp^2\,
J_0(\alpha\, q_\parallel)\, f(q_\parallel^2+q_\perp^2) & = &
\frac{4}{\alpha}\int_0^\infty dq\, q^2\,J_1(\alpha\, q)\, f(q^2)\ ,
\nonumber \\
\int_0^\infty dq_\parallel^2\ \int_0^\infty dq_\perp^2\;q_\perp^2\,
J_0(\alpha\, q_\parallel)\, f(q_\parallel^2+q_\perp^2) & = &
\frac{8}{\alpha^2}\int_0^\infty dq\, q^3\,J_2(\alpha\, q)\, f(q^2)\ .
\label{bessels}
\end{eqnarray}
Now, using Eqs.~(\ref{landausum}) and (\ref{bessels}) it is easy to see that
\begin{equation}
\hat X^{(p)}_{s_p}\Big|_{B=0} = \hat X\ ,\qquad\qquad \frac{\partial\hat
X^{(p)}_{s_p}}{\partial |B|}\Big|_{B=0} = (Q_p-Q_u)\,{\cal I}_1 - Q_p\,{\cal
I}_2\ ,
\end{equation}
where $\hat X$ and ${\cal I}_k$ are given by Eqs.~(\ref{AA}) and (\ref{ik}),
respectively.

A similar procedure can be followed in order to obtain the expansions for
$\hat Y^{(p)}_{s_p}$, $\hat X^{(n)}_\lambda$ and $\hat Y^{(n)}_\lambda$. The
evaluation of the derivatives in the denominator of Eq.~(\ref{alphaapp}) is
straightforward, leading to the final expressions of $\alpha_p$ and
$\alpha_n$ in Eqs.~(\ref{alphan}).

\end{document}